\documentclass[11pt,titlepage]{article}

\usepackage[round,numbers,sort&compress]{natbib} 

\usepackage{amsmath, amssymb}   

\usepackage{tabularx}

\usepackage{graphicx}

\usepackage[english]{babel}  
       
\usepackage{color}

\usepackage[scanall]{psfrag}

\title{\bf A Kinetic Model Describing the Processivity of Myosin-V}
 
\author{Karl I.~Skau$^{* \ddagger}$
	\and  Rebecca B.~Hoyle$^*$ \\%\thanks{Corresponding author.}
	 $^*$ Department of Mathematics and Statistics, \\
	University of Surrey, Guildford, Surrey, GU2 7XH, UK
	\and Matthew S.~Turner$^\ddagger$\\
	$^\ddagger$ Department of Physics,\\
	Warwick University, Coventry, CV4 7AL, UK
}

\date{}

\begin{document}  %%%%%%%%%%%%%%%%%

\maketitle

% --- --- --- --- section* : Abstract --- --- --- --- --- --- --- --- --- --- %
\abstract{The precise details of how myosin-V coordinates the
biochemical reactions and mechanical motions of its two head elements
to engineer effective processive molecular motion along actin
filaments remain unresolved. We compare a quantitative kinetic model
of the myosin-V walk, consisting of five basic states augmented by two further states to allow for futile hydrolysis and detachments, with experimental results for run lengths,
velocities, and dwell times and their dependence on 
bulk nucleotide concentrations and external loads in both directions. The model reveals how myosin-V can use the internal strain in the
molecule to synchronise the motion of the head elements. Estimates for the
rate constants in the reaction cycle and the internal strain energy are obtained by a computational comparison scheme involving an extensive exploration of the large parameter space. This scheme exploits the fact that we have obtained analytic results for our reaction network, e.g. for the velocity but also the run length, diffusion constant and fraction of backward steps. The agreement with experiment is often reasonable but some open problems are highlighted, in particular the inability of such a general model to reproduce the reported dependence of run length on ADP. The novel way that our approach explores parameter space means that any confirmed discrepancies should give new insights into the reaction network model. }

\clearpage

% --- --- --- --- section* : Introduction --- --- --- --- --- --- --- --- --- %
\section*{Introduction}

A myosin protein is an ATPase which gains enzymatic activity by
attaching to an actin filament \cite{Sellers_book, Howard_book,
Bray_book}. The myosin proteins use the chemical energy released in
ATP-hydrolysis to create directed mechanical motion. More than 100
proteins have been identified as belonging to the myosin super-family
and they are organised into some 18 subgroups \cite{Sellers_book,
Berg-Powell-Cheney2001}. A myosin is identified by a conserved 80 kDa
motor domain and it is usually assumed that all the myosin motor
proteins share the same biochemical reaction pathway when hydrolysing
ATP \cite{Sellers_book}.  The most studied of the myosins is the
non-processive (muscle) myosin II whose main reaction path is found to
follow the classical Lymn--Taylor scheme \cite{Lymn-Taylor1971}
describing the correlation of mechanical and chemical events.

Myosin-V is a dimeric protein involved in the intracellular transport
of a variety of cargos. The neck region of the two head elements is
three times the length of the corresponding region of the myosin II
heads
\cite{Cheney-OShea-Heuser-Coelho-Wolenski-Espreafico-Forscher-Larson-Mooseker1993}. 
It is generally assumed that the long
neck region acts as a lever arm and the size of the protein makes it
possible for myosin-V to walk hand-over-hand
\cite{Forkey-Quinlan-Shaw-Corrie-Goldman2003,
Yildiz-Forkey-McKinney-Ha-Goldman-Selvin2003,
Snyder-Sakamoto-Hammer-Sellers-Selvin2004,
Warshaw-Kennedy-Work-Krementsova-Beck-Trybus2005bpj} following the
helical repeat of actin.  Myosin-V was the first molecular motor shown
to be \emph{processive} along actin filaments
\cite{Mehta-Rock-Rief-Spudich-Mooseker-Cheney1999}. Previously
established processive motors like kinesin and dynein use microtubules
as the track for their processive directed motion, or in the case of
RNA polymerase, DNA. Naturally, there has been much interest both in
confirming the processive motion of myosin-V and in gaining insight
into the details of the molecule's motion using recent developments in
experimental techniques
\cite{Mehta-Rock-Rief-Spudich-Mooseker-Cheney1999,
Rief-Rock-Mehta-Mooseker-Cheney-Spudich2000, Mehta2001,
DeLaCruz-Wells-Rosenfeld-Ostap-Sweeney1999,
DeLaCruz-Sweeney-Ostap2000, DeLaCruz-Wells-Sweeney-Ostap2000,
Trybus-Krementsova-Freyzon1999, Sakamoto-Amitani-Yokota-Ando2000,
Wang-Chen-Arcucci-Harvey-Bowers-Xu-Hammer-Sellers2000,
Walker-Burgess-Sellers-Wang-Hammer-Trinick-Knight2000,
Nishizaka-Seo-Tadakuma-Kinosita-Ishiwata2000,
Veigel-Wang-Bartoo-Sellers-Molloy2002,
Forkey-Quinlan-Shaw-Corrie-Goldman2003,
Yildiz-Forkey-McKinney-Ha-Goldman-Selvin2003,
Warshaw-Kennedy-Work-Krementsova-Beck-Trybus2005bpj,
Snyder-Sakamoto-Hammer-Sellers-Selvin2004,
Baker-Krementsova-Kennedy-Armstrong-Trybus-Warshaw2004pnas,
Rosenfeld-Sweeney2004, Uemura-Higuchi-Olivares-DeLaCruz-Ishiwata2004,
Clemen-Vilfan-Jaud-Zhang-Barmann-Rief2005}. Similarly, several
theoretical models of the myosin-V walk with different levels of
detail have been proposed
\cite{Rief-Rock-Mehta-Mooseker-Cheney-Spudich2000,
Walker-Burgess-Sellers-Wang-Hammer-Trinick-Knight2000, Mehta2001,
Veigel-Wang-Bartoo-Sellers-Molloy2002, Kolomeisky-Fisher2003bpj,
Baker-Krementsova-Kennedy-Armstrong-Trybus-Warshaw2004pnas,
DeLaCruz-Ostap2004, Vale2003,
Uemura-Higuchi-Olivares-DeLaCruz-Ishiwata2004, Rosenfeld-Sweeney2004,
Vilfan2005, Lan-Sum2005}.

How myosin-V coordinates the biochemical reactions and mechanical
motions of the two head elements of the protein to become an effective
processive molecular motor is an open question.  While there is a
general agreement on the enzymatic reaction path for a \emph{single}
myosin head, the details of how myosin-V keeps the two heads' reaction
cycles in phase is still unresolved. Likewise, there is still a lack
of understanding of how external forces directly influence the kinetic
mechanism of the walk and what role the internal strain in the
molecule plays. As a result of the large number of measurements made
on myosin-V in the last years, this is an appropriate time to make
more detailed, quantitative models of the myosin-V walk, and to see
how such models compare with what is found experimentally.

The outline of the article is as follows: we first establish our model
for the processive walk of myosin-V and explain the model's underlying
assumptions.  In the results section we present analytic results for our, rather general, reaction network. These include the velocity, run length, diffusion constant and fraction of backward steps. These observable quantities are functions that depend on a number of parameters (mainly characteristic energies for the various transitions between states). The unknown parameters
are estimated by optimising the agreement of the model with a chosen
representative set of experimental results. This approach, which is
based on defining a cost-function, is outlined in the third section. We
then present quantitative results for the optimised parameters and
compare them with available experimental data and end with a
discussion. Agreement with experiment is generally reasonable, though we are unable to explain the trend of the run length with varying ADP concentration. We discuss how our approach might be used to explore such possible discrepancies between existing reaction network models and experiment, generating new insights into the myosin V stepping mechanism.

% --- --- --- --- section* : the Model    --- --- --- --- --- --- --- --- --- %
\section*{Model}

The reaction network for our model is sketched in
Fig.~\ref{fig:ModelFig}. The main reaction cycle contains states 1 to
5. In state 4 the rear head starts its reaction cycle by releasing
ADP, while the front head will not react before the whole cycle has
been completed making it the new rear head. In state 5 the rear head
is in the rigor state with no nucleotide attached, and the rear head
will detach rapidly from the actin filament when it reacts with ATP
from the bulk. The mechanical motion that moves the rear head into the
front position is found experimentally to happen in two steps
\cite{Veigel-Wang-Bartoo-Sellers-Molloy2002,
Uemura-Higuchi-Olivares-DeLaCruz-Ishiwata2004}, from position 1 to
position 2, and then from position 2 to position 3. In state 3 the
front head is weakly attached to actin, so only one head is strongly
attached in states 1 to 3, making these the more vulnerable states for
total detachment of the molecule from actin. When inorganic phosphate
${\rm P_i}$ is released from the lead head, going from state 3 to
state 4, the myosin binds strongly to actin and the lead head makes
the so-called ``powerstroke''.  Since the rear head is still attached,
the lead head will not achieve the usual post-powerstroke angle with
the actin filament, which will cause internal strain in the
molecule. This is consistent with electron microscopy (EM) images
\cite{Walker-Burgess-Sellers-Wang-Hammer-Trinick-Knight2000,
Snyder-Sakamoto-Hammer-Sellers-Selvin2004} which show strongly bound
myosin-V to take a position similar to a telemark-skier's stance,
indicating large internal strain in the molecule.

The model given above for the main walk follows fairly closely the
intermediate states proposed by
\citet{Rief-Rock-Mehta-Mooseker-Cheney-Spudich2000} which accounts for
the most important experimental findings for myosin-V
\cite{Mehta2001}. Note that there is still disagreement in the
literature as to which states are present in the main reaction cycle
for the processive motion of myosin-V (see for instance
Refs.~\cite{Mehta2001, Veigel-Wang-Bartoo-Sellers-Molloy2002,
Baker-Krementsova-Kennedy-Armstrong-Trybus-Warshaw2004pnas,
DeLaCruz-Ostap2004, Rosenfeld-Sweeney2004, Vilfan2005}). A natural consequence of the microscopic size of the motor protein is that both mechanical and chemical effects are important. We expect the external force, due for instance to the viscous
drag of a cargo vesicle, to most strongly influence the mechanical
steps in the cycle which involve translational motion.

The main reaction path is reversible in the sense that one could, in principle, run the ATP
hydrolysis reaction in reverse by pulling the motor backwards with an
external force. (This would not be a very effective way to produce ATP
since the motor is kinetically tuned to move efficiently in one
direction only). It is important for a model which aims to investigate
the mechanism of a motor protein to also include the effect of futile
cycles and detachment rates. Futile cycles consume ATP without
creating any net movement of the molecule
while detachment limits the run length of
the molecule. If, in an evolutionary sense, myosin-V were to tune the reaction rates of the motor
domain it might well seek to maximise the forward velocity (requiring ``weak binding"). However, it should probably
also seek to minimise the impact of futile cycles and detachment rates (requiring ``strong binding"). The kinetic parameters that arise might then be expected to correspond to a suitable compromise solution. We remark that the computational scheme for exploring parameter space that we will later describe could be used in the future to explore such evolutionary pressures and trends.

There will in general be a large number of possible unfavourable
pathways, but we will assume that among these there is one dominant
futile reaction cycle which takes place when the ADP in state 4 of the
main cycle (Fig.~\ref{fig:ModelFig}) detaches from the front head
before the rear head moving to state 6. In the futile cycle the front
head then reacts with a new ATP molecule so that the molecule returns
back to state 2 without having created any net movement. Similarly, we
will identify what we believe to be the dominant detachment rate. We
make the conjecture that the molecule is most vulnerable to detachment when only
one head is attached to actin, as in state 2 (due to the high internal
strain in the molecule, state 1 is a short lived unstable state).
We will assume that detachment from state 2 dominates over all other
detachment rates.  The mechanism of detachment will be the release of
ADP followed by the binding of ATP to the rear head so that
myosin-V detaches completely from the actin filament
before strong
attachment is achieved by the front head (state 7, Fig.~\ref{fig:ModelFig}).

\begin{figure}
  \centering
   \includegraphics[width=100mm]{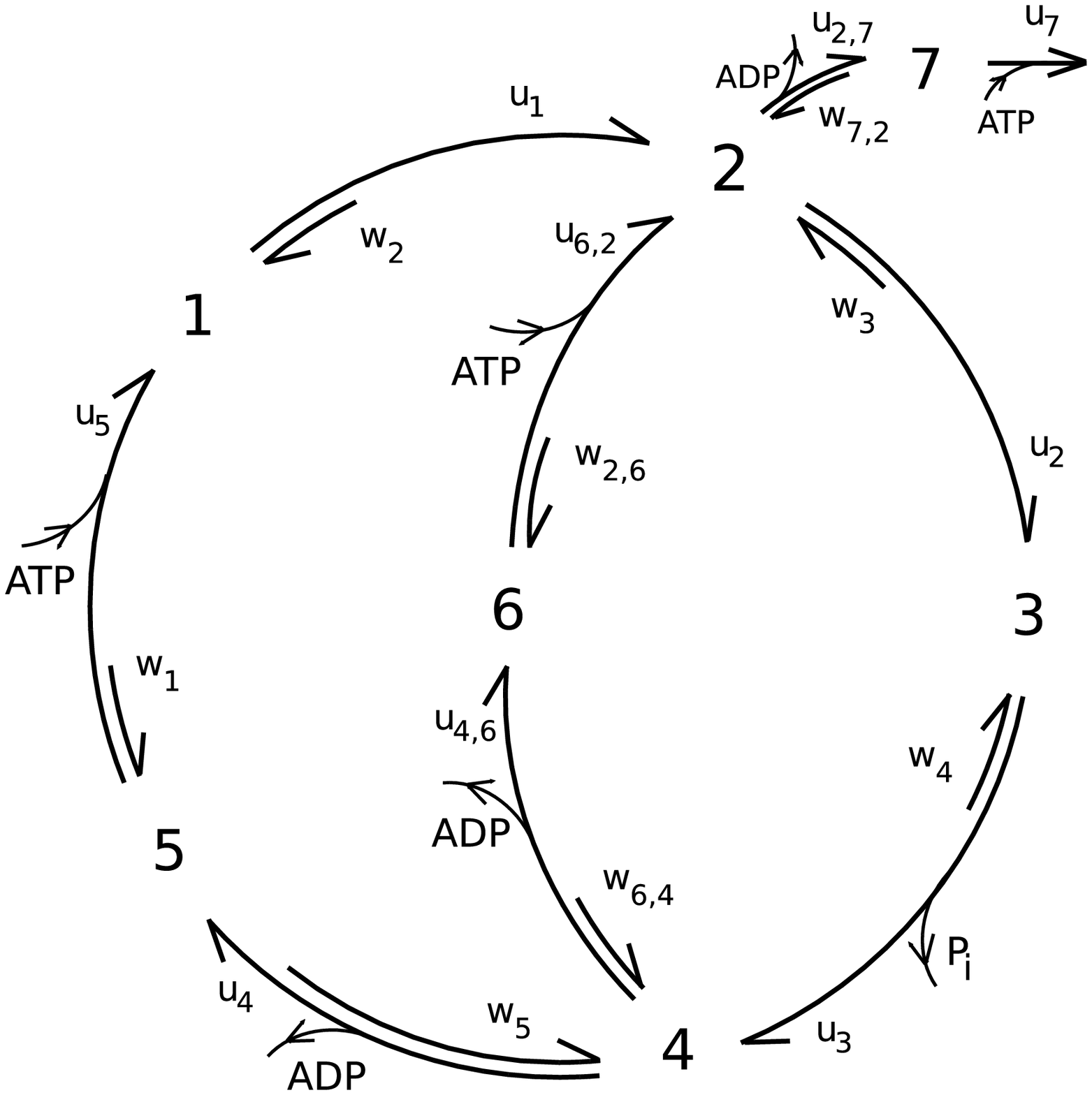}
   \caption{}
        \label{fig:Cycle}
\end{figure}

The assumption we have made so far is that the main cycle is described
well by the qualitative model of
\citet{Rief-Rock-Mehta-Mooseker-Cheney-Spudich2000} where in addition
the mechanical motion of the rear head to the front is taken in two
steps. Furthermore, we assume that there is one futile cycle that
dominates over all others, and that this futile cycle involves states
2 and 4 in the main cycle. Lastly, we assume that the detachment rate
at state 2 dominates all others. The complete model we have outlined
(shown in Figs.~\ref{fig:ModelFig} and \ref{fig:Cycle}) is a minimal
realistic model for the walk of a molecular motor like myosin-V.

% --- --- --- --- 
\subsection*{Reaction rates}

The reaction rates between the different states are described by
Arrhenius expressions:
\begin{equation}
  w_{i} = \tau^{-1} \,\,
             {\rm e}^{-(G_{i}^{\ddagger}\, + \, \Delta G_{i})/k_{\rm B}T} \,,
\end{equation}
where $\tau$ is the fundamental time scale of the reaction and $k_{\rm
B}$ is the Boltzmann constant. (We use the notation where $u_i$ and
$w_i$ are the forward and backward rates, respectively, away from
state $i$ \cite{Fisher-Kolomeisky1999pA}). $G_{i}^{\ddagger}$ is the
energy barrier between state $i$ and its neighbour state in the
forward direction, while $\Delta G_{i}$ is the energy difference
between the two states. We set $T = 298$~K in this article.
The total energy balance for the ATP hydrolysis is
\begin{equation}
  \Delta G_{\rm hyd} = 
       k_{\rm B}T\ln\biggl(\frac{\rm [ATP]}{\rm [ADP][P_i]}\biggr) 
         + \sum\limits_{i=2}^{5} \Delta G_{i}  \,,
   \label{Eq:Ghyd}
\end{equation}
where the nucleotide concentrations are made dimensionless by dividing
by the concentrations at the reference states for $\Delta G_{i}$,
[ATP$]^0$=[ADP$]^0$=[${\rm P_{i}}]^0$ = 1 M. 
The standard free energy $\Delta G^{(0)} =  \sum \Delta G_{i} \simeq 32.5\, [k\mathrm{J/mol}]$
\cite{Alberty-Goldberg1992} which is close to 13 $k_{\rm B} T$, while 
$\Delta G_{\rm hyd} \simeq$  25 $k_{\rm B} T$ at cellular conditions \cite{Howard_book}.

\begin{figure}
  \centering
   \includegraphics[width=100mm]{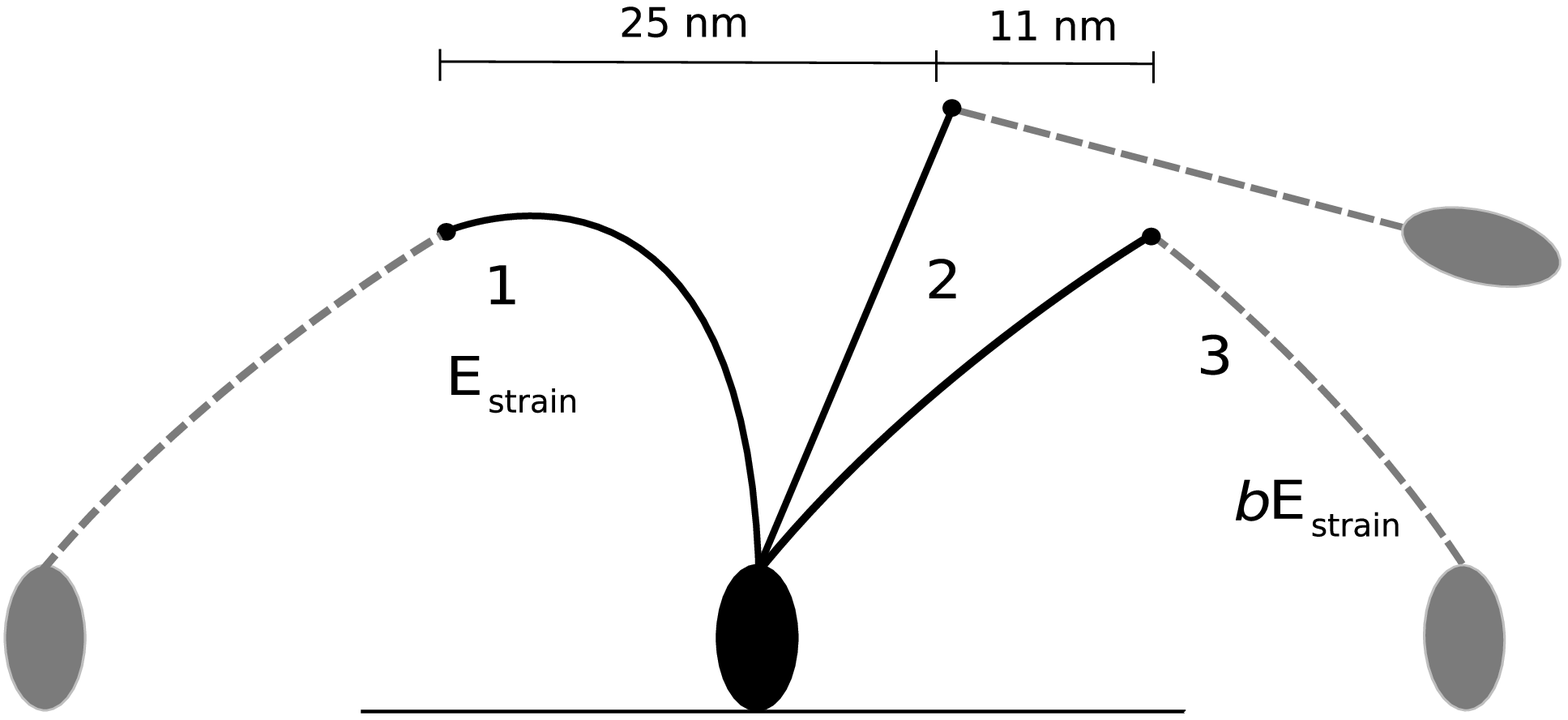}
   \caption{}
        \label{fig:Mech}
\end{figure}

For the main cycle, one ends up with a total of ten reaction rates.
The first four rate constants in the main cycle are related to the
mechanical movement(see Figs.~\ref{fig:ModelFig} and \ref{fig:Mech});
\begin{eqnarray}
	 %%%%
  u_{1}   &=& \tau_{\rm d}^{-1} \,,
                     \\
  w_{2}   &=& \tau_{\rm d}^{-1} \,\,
             {\rm e}^{-(E_{\rm strain}\, -\, f_{\rm ex}\, (d_{W} - \frac{1}{2} \, f_{\rm ex}/ k_{\rm H})\,)/k_{\rm B}T} \,,
          \label{eq:w[2]} \\ 
	 %%%%
  u_{2} &=& \tau_{\rm d}^{-1} \,\, 
       {\rm e}^{-(G_{2}^{\ddagger}\, 
	   +\, f_{\rm ex}\, (d_{D} + \frac{1}{2} \, f_{\rm ex}/ k_{\rm H})\, + \, b E_{\rm strain}) /k_{\rm B}T} \,,
           \label{eq:u[2]}    \\
  w_{3} &=& \tau_{\rm d}^{-1} \,\,
                 {\rm e}^{-(G_{2}^{\ddagger}\, +\, \Delta G_{2})/k_{\rm B}T} \,, 
	 %%%%     
\end{eqnarray}
where $\tau_{\rm d}$ is a hydrodynamic time-scale related to diffusion
over one step-length and $f_{\rm ex}$ is the component of the external force parallel to the actin filament. The mechanical step is 
separated into a
so-called working stroke of $d_{W} \simeq 25$~nm and a diffusional
sub-step of $d_{D} \simeq 11$~nm (see Fig.~\ref{fig:Mech}),
as indicated by experimental findings
\cite{Veigel-Wang-Bartoo-Sellers-Molloy2002, Uemura-Higuchi-Olivares-DeLaCruz-Ishiwata2004}, giving a total step size
$d = d_{W} + d_{D} \simeq 36$~nm
\cite{Mehta-Rock-Rief-Spudich-Mooseker-Cheney1999,
Rief-Rock-Mehta-Mooseker-Cheney-Spudich2000,
	Sakamoto-Amitani-Yokota-Ando2000}. Thus we neglect the (weak) external force dependence of the diffusional step size in $u_1$ but capture the dominant effect of the force in, e.g. retarding the activated rate $u_2$ and in distorting the metastable shape of the bound arm in state 2 (through the $\sim f_{\rm ex}^2$ terms in the exponents of $w_{2}$ and $u_{2}$). The total internal strain in the
molecule $E_{\rm strain}$ defines an effective Hookeian spring
constant, $k_{\rm H}$, related to the compliance in the motor and neck
region of the myosin head:
\begin{equation}
  E_{\rm strain} = \frac{1}{2}\, k_{\rm H}\, d_{W}^2 \,,
    \label{eq:Hookeian_spring}
\end{equation}
where we have assumed that the molecule is fully strained in state 1.

State 2 is in mechanical equilibrium. The position of the hinge/neck
in state 2 is influenced by the magnitude of the external force, which
gives rise to the energy term $1/2 \, f_{\rm ex}^2/ k_{\rm H}$ in $
w_{2}$ and $u_{2}$ (Eqs.~\eqref{eq:w[2]} and \eqref{eq:u[2]}).  When
moving from state 2, either back to state 1 or forward to state 3, the
molecule increases its internal strain by $E_{\rm strain}$ or $b
E_{\rm strain} = 1/2\,k_{\rm H}\, d_{D}^2$ respectively
(Fig.~\ref{fig:Mech}). Using the latter expression together
with Eq.~\eqref{eq:Hookeian_spring} gives $b = (d_{D}/d_{W})^2
\simeq 0.2$. Since $E_{\rm strain} > b E_{\rm strain}$ there
is a bias in the forward direction away from state 2, making the
molecule a Brownian ratchet \cite{Julicher-Ajdari-Prost1997}. 

Notice that the external force $f_{\rm ex}$ is defined to be positive
in the direction opposite to the movement of the myosin-V molecule,
and that only the influence of the force parallel to the actin
filament is taken into account. (The external force is of course a
vector quantity, with a possible important influence on the walk
arising from the normal component of the force \cite{Fisher-Kim, Kim-Fisher}.)
The external force will accelerate or slow down the two reaction
rates $w_{2}$ and $u_{2}$ away from state 2, depending on the sign of
$f_{\rm ex}$.

Experimentally, it is found that there is one rate-limiting step in
the walk of myosin-V \cite{DeLaCruz-Wells-Rosenfeld-Ostap-Sweeney1999,
DeLaCruz-Sweeney-Ostap2000, DeLaCruz-Wells-Sweeney-Ostap2000}. At
large external force, $f_{ex}$, one or more sub-steps that couple to
the force, will become rate-limiting. A phenomenological way to model
this is \cite{Mehta-Rock-Rief-Spudich-Mooseker-Cheney1999}
\begin{equation}
  \tau_{*} = \tau_1 + \tau_2\, {\rm e}^{f_{ex} d_{\rm eff}/ k_{\rm B}T} \,.
    \label{eq:phenomenological}
\end{equation}
Here $\tau_{*}$ is the (average) dwell time for one step, while
$\tau_1$ and $\tau_2$ are the dwell times of the force independent and
dependent sub-steps respectively, and the external force couples to a
distance $d_{\rm eff}$. When fitting experimental data to this
phenomenological equation, naively one would expect $d_{\rm eff}
\simeq 36$ nm which is the average step length. Instead it was
found that $d_{\rm eff}$ is between 10--15 nm.  From our model
this scale emerges quite naturally, since the external force
couples to the diffusional search governed by rate $u_2$ which gives
$\tau_{*} \simeq \tau_1 + 1/u_2$. From Eq.~\eqref{eq:u[2]} we have
\begin{equation}
  u_{2}  = A \, {\rm e}^{-f_{\rm ex}\, (d_{D} + \frac{1}{2} \, f_{\rm ex}/ k_{\rm H})/ k_{\rm B}T} \,,
\end{equation}
where $A$ is some constant.  Here $d_{D}$ is 11 nm while the
correction term $\frac{1}{2} \, f_{\rm ex}/ k_{\rm H}$ of the position
of the hinge in state 2 is of the order of 8 nm when close to stall
force. It is satisfying that this is consistent with the value of $d_{\rm eff}$
cited above and indicates that the way we include the two substeps in our model is reasonable. 
(See Ref.~\cite{Kolomeisky-Fisher2003bpj} for a discussion of the limitations of
Eq.~\eqref{eq:phenomenological}).

The reaction rates for states with both heads attached to actin (see
Fig.~\ref{fig:ModelFig}) are given by
\begin{eqnarray}
	 %%%%
  u_{3} &=& \tau^{-1} \,\,
          {\rm e}^{-G_{3}^{\ddagger}/k_{\rm B}T} \,,
                    \label{eq:u3}\\
  w_{4} &=& [{\rm P_{i}}] \,\, \tau^{-1} \,\,
                 {\rm e}^{-(G_{3}^{\ddagger}\, +\, \Delta G_{3}\,
                    - \, (1 - b) E_{\rm strain})/k_{\rm B}T} \,,
		    \label{eq:w4}
           \\
	 %%%%
  u_{4}   &=& \tau^{-1} \,\,
                 {\rm e}^{-G_{4}^{\ddagger}/k_{\rm B}T} \,,
                    \\
  w_{5}   &=& [{\rm ADP}]\,\, \tau^{-1} \,\,
                 {\rm e}^{-(G_{4}^{\ddagger}\, +\,\Delta G_{4})/k_{\rm B}T} \,,
            \\ 
	 %%%%
  u_{5}    &=& [{\rm ATP}]\,\, 
                 \tau^{-1} \,\, {\rm e}^{-G_{5}^{\ddagger}/k_{\rm B}T} \,,
                        \label{eq:maincycle_rate}
	             \\ 
  w_{1}    &=& \tau^{-1} \,\, 
                 {\rm e}^{-(G_{5}^{\ddagger}\, +\, \Delta G_{5})/k_{\rm B}T}
		 \label{eq:w1}
               \,,  
	 %%%%     
\end{eqnarray}
where the nucleotide concentrations and
[${\rm P_i}$] in Eqs.~\eqref{eq:w4}--\eqref{eq:maincycle_rate} are given as 
dimensionless quantities 
(see comments under Eq.~\eqref{Eq:Ghyd}). $\tau$ is a microscopic time scale related to the characteristic
oscillation frequency of the protein. The free parameters in the model are
the activation energies $G_{i}^{\ddagger}$, the energy differences $\Delta G_i$, and
the strain energies. When an estimate is made for these free parameters, the
resulting energy landscape (Fig.~\ref{fig:EnergyLand}) gives directly all the predicted
reaction rates, and the model's predictions for myosin-V's velocity and
run lengths along the actin filament (see Appendix~\ref{App:Expression}). 
Note that $\tau$ is \emph{not} really an independent variable
since it can be absorbed into the activation energies; 
$1/\tau\, \exp(G^\ddagger_i) \equiv 1/ \tau_0\, \exp(G^\ddagger_i + \ln(\tau_0/\tau))$.

As a measurement of the deviation away from equilibrium one can
introduce the parameter
\begin{equation}
  \Gamma = \prod_{j=1}^{5}  \frac{u_j}{w_j}
         = {\rm e}^{\Delta G_{\rm hyd}/ k_{\rm B} T}
        \,{\rm e}^{-f_{\rm ex}d/ k_{\rm B} T} \,,
    \label{eq:Gamma}
\end{equation}
which gives the thermodynamic driving force for the molecular motor
\cite{Onsager1931prI, Fisher-Kolomeisky1999pA}. To fulfill detailed
balance we have $\Gamma = 1$ at equilibrium \cite{Onsager1931prI}. As
expected from the energy balance the external force appears in
$\Gamma$ as $f_{\rm ex} d$; this is the total work done by the motor
when completing one reaction cycle with step-length $d$. From
Eq.~\eqref{eq:Gamma} it is clear how the external force shifts the
apparent equilibrium constant of the hydrolysis reaction . This
coupling between the external force and the free energy of the
hydrolysis of ATP, gives directly the thermodynamic upper bound on the
stall-force $f_{\rm stall}=\Delta G_{\rm hyd}/d$ ($\simeq 2.8$~pN at
cellular conditions). (The presence of futile cycles will influence the
upper bound on the stall-force, but this correction is found to be insignificant
for our (best) model).

The reaction rates for the futile cycle are given by
\begin{eqnarray}
	 %%%%     
  u_{4,6}   &=& u_{4} \,\,
        {\rm e}^{-\alpha E_{\rm strain}/k_{\rm B}T}  \,,
	\label{eq:u46}
                    \\
  w_{6,4}   &=& w_{5} \,\,
                 {\rm e}^{-\alpha E_{\rm strain} /k_{\rm B}T}  \,,
	\label{eq:w64}
            \\ 
	 %%%%
  u_{6,2}   &=& u_{5}   \,,
		 \label{eq:futilecycle_rate}
                    \\
  w_{2,6}   &=&  w_{1} \,\, {\rm e}^{- E_{\rm strain} /k_{\rm B}T}  \,
  {\rm e}^{ -f_{\rm ex}\, (d_{D} + \frac{1}{2} \, f_{\rm ex}/ k_{\rm H})/k_{\rm B}T} \,. 
                \label{eq:futile_back}
	 %%%%
\end{eqnarray}
Here $u_{i, j}$ and $w_{i, j}$ are the reaction rates from state $i$
to state $j$. Using the fact that the two head elements are
structurally identical, we make the assumption that the front head has
to have the neck at a similar angle to that between the rear neck and
the actin filament in step 4 of the main cycle to make it possible for
ADP to be released or bind to the myosin motor domain
(Eqs.~\eqref{eq:u46} and \eqref{eq:w64}). To achieve this, the front
head has to overcome the strain energy in the molecule in state 4
which creates an energy barrier $\alpha E_{\rm strain}$ for the ADP
release reaction governed by rate $u_{4,6}$ (see Fig.~\ref{fig:Mech}). In this way myosin-V
synchronises the biochemical reactions of the two heads.

To move from state 2 to state 6 (Fig.~\ref{fig:ModelFig}), the
molecule has to increase the internal strain by $E_{\rm strain}$
without the energy from the hydrolysis reaction,
(Eq.~\eqref{eq:futile_back}), so there is a very low probability for
the futile cycle to run in reverse (effectively becoming a useful
reaction path way).  The reaction rates in the futile cycle follow the
reaction rates for the main cycle except for the extra barrier caused
by the strain in the molecule, so there is only one new parameter that
appears in the the model. In principle $\alpha E_{\rm strain}$ could
be estimated if the elastic moduli of the different parts of the
molecule were known, by making a detailed structural model based on EM
measurements
\cite{Walker-Burgess-Sellers-Wang-Hammer-Trinick-Knight2000} and
crystal structures \cite{Geeves-Holmes1999, Houdusse-Sweeney2001} of
myosin-V (see also \citet{Vilfan2005}). We will not attempt to do this here, but leave $\alpha
E_{\rm strain}$ as an undetermined energy barrier, an energy barrier
used by myosin-V to synchronise the reaction cycles of the two head
elements.

The detachment from state 2 takes place if the ADP detaches from the
rear head before the front head becomes weakly bound. The relevant
reaction rates are
\begin{eqnarray}
	 %%%%
  u_{2,7}   &=& u_{4} \,,
                     \\
  w_{7,2}   &=& w_{5} \,,
           \\ 
	 %%%%
  u_{7} &=& u_{5}\, {\rm e}^{ \, |f_{\rm ex}| \delta/k_{\rm B}T}  \,.
	       \label{eq:detachment_rate}
	 %%%%
\end{eqnarray}
The external force will increase the detachment rate of the single
head (an effect we neglect when both heads are strongly attached,
which should be a good approximation when considering the strain level
in the molecule). Pulling experiments on S1 give an apparent
interaction distance of $\delta = 2.4$~nm
\cite{Nishizaka-Seo-Tadakuma-Kinosita-Ishiwata2000} for the external
force (Eq.~\eqref{eq:detachment_rate}).

An extended model where we would also consider the reattachment rate
of motors is of course possible
\cite{Nieuwenhuizen-Klumpp-Lipowsky2004} but not particularly relevant
since the local bulk concentration of motors and of actin target sites
is usually not well-controlled in an experiment. Our main focus in
this model will be single-molecule experiments and their predictions.

In our model we use the fact that the two head regions of the myosin-V
protein are identical and have identical biochemical reaction
paths. It might seem at first that this would be an obstacle to the
processive motion of the protein, since the reactions of the two heads
must be out of phase to ensure that at all times at least one of the
two heads is strongly attached to the actin filament. However,
evidence has been found in EM experiments that the intra-molecular
strain affects the two bound heads asymmetrically
\cite{Walker-Burgess-Sellers-Wang-Hammer-Trinick-Knight2000,
Snyder-Sakamoto-Hammer-Sellers-Selvin2004}. The model outlined above
shows explicitly how this asymmetrical strain can be used by the two
heads to coordinate their reaction cycles and to minimise the impact
of the futile cycles and detachment rates.

% --- --- --- --- 
\subsection*{Parameters}

The fundamental time scales, $\tau$ and $\tau_{\rm d}$ can be
estimated using the Stokes--Einstein relation. $\tau_{\rm d}$ is
related to the diffusion time of the whole head element (with a
``diameter'' of $\simeq 30$ nm) over the step size $d_{W} \simeq
25$ nm, which gives $\tau_{\rm d} \simeq 10^{-5}$ s. Similarly,
$\tau$ is related to movement of a few nanometres, where the relevant
length scale is the thickness of the head element, giving $\tau
\simeq 10^{-8}$ s.  Note that any difference between $\tau$ or
$\tau_{\rm d}$ in the different steps will be absorbed into the
activation energies $G^{\ddagger}_i$ (see comment under
Eq.~\ref{eq:w1}) so we are only interested in the
order of magnitude of these time scales. Similarly, the activation
energies will also be modified by the bulk
pH and ionic strength, even though the influence of these solution
properties are not included explicitly into the model.

Estimates of some of the activation energies are available 
through chemical kinetic reaction rate measurements, but we will not try
to guess the values of the activation energies but leave them all as free 
parameters.
The model of the myosin-V walk outlined above is a fairly detailed model 
containing seven different states and 13 reaction rates, still the number of
undetermined parameters is relatively low. We have the four activation
energies, $G^{\ddagger}_i$, three independent energy differences $\Delta G_i$ (where one energy
difference will be dependent because of Eq.~\eqref{Eq:Ghyd}),
and finally two terms 
connected to the strain level in the molecule, $E_{\rm strain}$ and
$\alpha E_{\rm strain}$.  We are therefore left with nine undetermined
parameters in our model of the myosin-V walk, which is a small number
considering that the model is able to predict not only the walk's
dependence on the bulk concentrations of ATP, ADP, and ${\rm P_i}$,
but also how the external force couples to the walk. The model is also
detailed in its prediction of how the internal strain influences the
walk and the level of strain in the molecule. Our model is therefore
over-determined with respect to the available experimental data and
provides many measurable predictions.

% --- --- --- --- section* : Optimisation --- --- --- --- --- --- --- --- --- %
\section*{Optimisation}
  \label{sec:Opt}

To search the parameter space computationally we define a \emph{cost
function} which quantifies the agreement between the model and the
data. There is clearly some subjectivity involved in choosing the terms in the cost function: for example, one might be more inclined to include data that have been confirmed by several different research groups. We cannot investigate all possible permutations of our cost function since the optimisation approach is quite computationally demanding. However, the 17 experimental features that we identify, and encode in
the cost function, are those that we expect a good model for the walk of
myosin-V to reproduce. These include trends observed under
variation of nucleotide bulk concentrations and the external
force. Each term in the cost function was included to search for such a trend or to restrict the value of the energy jumps $\Delta G_i$ between substeps. Likewise, we specify variances to indicate how large a deviation
from each target value we deem acceptable. The cost function,
$\Delta$, given in Appendix~\ref{App:CostFun}, is constructed as a simple sum of 17 terms and each term gives an $O(1)$ contribution when within the accepted experimental error (as defined by our chosen variance). We believe that our cost function is a minimal encoding of the most important experimental trends, but it is possible to extend or modify our approach by including further or different data points if desired.

We do not choose arbitrarily what values the nine undetermined
parameters should take but instead search the parameter space for
favourable combinations using a technique based on simulated annealing
\cite{Galassi-Davies-Theiler-Gough-Jungman-Booth-Rossi_book} of the
cost function. To make the search effective, we first evaluate the
cost function at 50 million points in a parameter space of nine times
25 $k_{\rm B}T$. (This is the energy available in the hydrolysis 
reaction, Eq.~\eqref{Eq:Ghyd}, under cellular conditions. The internal
energy $E_{\rm strain}$ and energy differences $\Delta G_i$ cannot be 
larger than the energy in the
hydrolysis reaction, while there are no such limits in principle on
the kinetic parameters and $\alpha E_{\rm strain}$). To make sure
these points are evenly spread in the parameter space we use the Sobol
quasi-random sequence \cite{Locke-Millar-Turner2005,
Galassi-Davies-Theiler-Gough-Jungman-Booth-Rossi_book}. The fifty
points with the lowest cost function from the Sobol sampling were then
passed to a simulated annealing routine, where the cost function is
the ``energy'' term. The number of steps in the simulated annealing is
chosen so that it is equal to the number of steps in a random walk
over the average distance between the Sobol points.

An attractive feature of estimating the free parameters in the model
like this is that it has some similarities to the way the molecular motor
has tuned the same parameters through evolution to achieve
physiologically required velocities and run lengths under variable
cellular conditions.

% --- --- --- --- section* : Results      --- --- --- --- --- --- --- --- --- %
\section*{Results}

\begin{figure}
  \centering
   \includegraphics[width=150mm]{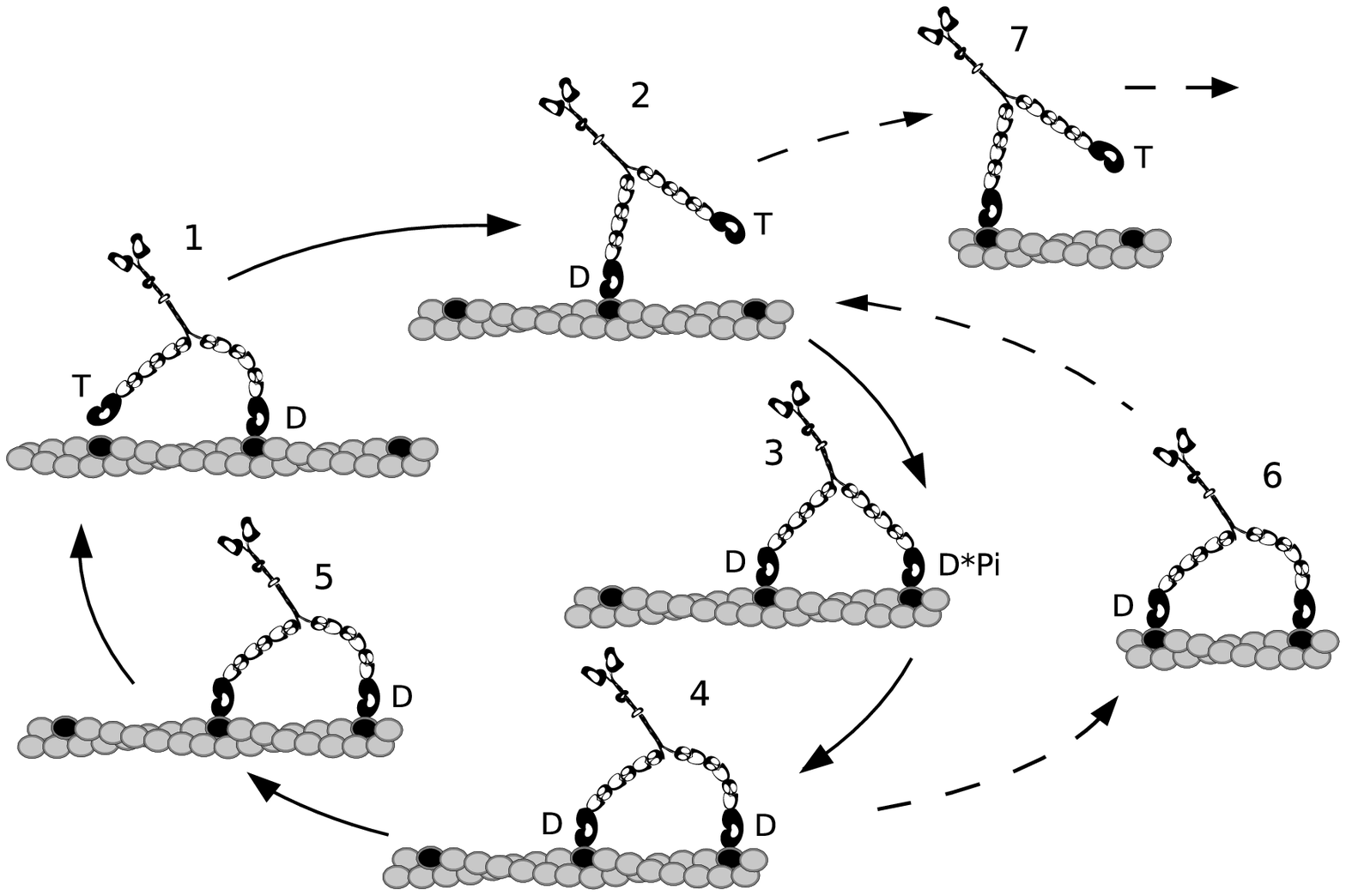}
   \caption{}
        \label{fig:ModelFig}
\end{figure}

After optimising the parameters for our model (Table~\ref{tab:parameters}),
it was found that there is one
rate-limiting step, $u_4=14.7$ $s^{-1}$, corresponding
to the largest value of $G_{i}^\ddagger$ for $i=2,...,5$,
which is the release of ADP from the rear head in state 4,
Fig.~\ref{fig:ModelFig}. The ADP release rate is in quantitative
agreement with kinetic measurements \cite{Mehta2001,
DeLaCruz-Wells-Rosenfeld-Ostap-Sweeney1999,
DeLaCruz-Sweeney-Ostap2000, DeLaCruz-Wells-Sweeney-Ostap2000,
Mehta-Rock-Rief-Spudich-Mooseker-Cheney1999,
Rosenfeld-Sweeney2004,Veigel-Wang-Bartoo-Sellers-Molloy2002} which 
estimates $u_4$ between 10 $s^{-1}$ to 20 $s^{-1}$. From the
model (Fig.~\ref{fig:ModelFig}) it can be seen why ADP release is
the crucial reaction step for the myosin-V walk, since both the futile cycle
and detachment depend on it. By slowing down
the ADP release, myosin-V achieves a larger duty ratio for the myosin
head, but more importantly, also reduces the flux around the
futile cycle and similarly the detachment rate. Since the rate
limiting step also determines the average velocity of the motor, there
is a trade off when tuning the rate of $u_4$. It is satisfying that
the optimisation scheme of our model is able to reproduce the kinetic
tuning found experimentally. It is worth noting that no optimal
solutions were found in other parts of the parameter space, i.e., no
other possible combination of reaction rates could be found with
either a different rate-limiting step or more than one rate-limiting
step.

The optimised ``best fit" model parameters, as given in Table~\ref{tab:parameters}, 
give rise to $u_5$ = 0.3 $\mu$M $s^{-1}$ for the attachment rate
of ATP to the rear head in state 5, a prediction that is somewhat slower than 
experimental estimates of $u_5$ in the range
from 0.6--1.5 $\mu$M $s^{-1}$ \cite{Veigel-Wang-Bartoo-Sellers-Molloy2002}.  The
release of ${\rm P_i}$ is found to be fast when the front head is
attached weakly to actin (state 3, Fig.~\ref{fig:ModelFig}) with
$u_3$ = 3200 $s^{-1}$, an order of magnitude above the lower bound from kinetic measurements
$u_3 > 250 s^{-1}$ \cite{DeLaCruz-Wells-Rosenfeld-Ostap-Sweeney1999}.

It is a somewhat subtle point that the velocities and run lengths that our model predicts (see Appendix~\ref{App:Expression}) are independent of one aspect of that model, specifically whether the mechanical motion between state 3 and state 4 is assumed to happen before or during the release of ${\rm P_i}$. This is in spite of the fact that the forms of Eqs.~\eqref{eq:u3} and \eqref{eq:w4} {\em are} sensitive to this difference: the $(1-b) E_{\rm strain}$ term would appear in the exponential in $u_3$ and not  in $w_4$ if the mechanical transition occurred before the phosphate release rather than during the phosphate release, as we assume here. The reason for this is that the optimization process fixes only the observable rates ($u_3$ and $w_4$). This appears to suggest that there may be some ambiguity in the value of the energy $G^{\ddagger}_3$ that appears in these rates, depending on whether the motion between state 3 and state 4 happens before or during the release of ${\rm P_i}$, i.e. whether the peak in the energy landscape occurs closer to state 4 or state 3 respectively. We can nonetheless determine the form of the equations by considering the kinetic measurements for ${\rm P_i}$ release. These are carried out using single myosin heads \emph{without} internal strain. If, for the intact two-headed molecule, the $(1-b) E_{\rm strain}$ term was instead placed in the exponential in $u_3$ (and the peak in the energy landscape was close to state 4) then the rate for a {\em strain-free} transition, such as would be expected to be the case for single headed molecules, would be larger by a factor of $\exp[(1-b)E_{\rm strain}]$, giving rise to unrealistic phosphate release rates that would be four orders of magnitude higher than  3200 $s^{-1}$. This provides quantitative evidence that the power stroke of the front lever arm takes place substantially \emph{after} ${\rm P_i}$ is released, as is illustrated in Fig.~\ref{fig:EnergyLand} where the energy barrier between state 3 and state 4 is close to state 3.

We estimate the internal strain, $E_{\rm strain}=12.8$ $k_{\rm B}T$ which gives a rigidity $\kappa$ = 120 $k_{\rm B}T$ nm = 500 pN
nm$^2$  for the lever arm \cite{Landau-Lifshitz_elasticbook}, in
agreement with other estimates in the literature
\cite{Howard_Spudich1996}. This gives a Young modulus of $Y$ = 0.6
[GPa] assuming the myosin neck has an effective radius of 1 nm, which
is comparable to what is found in similar proteins \cite{Howard_book}.
The strain barrier, $\alpha E_{\rm strain}$, preventing the futile cycle
was found in the optimisation scheme to be 5.4 $k_{\rm B}T$, which slows
down the release of ADP from the front head (state 4, Fig.~\ref{fig:ModelFig})
$\sim$ 50--200 times compared to the rear head under changing nucleotide concentrations
in the bulk. This is consistent with measurements reported by \citet{Rosenfeld-Sweeney2004}. 

Among the more important properties of myosin-V are its velocity and
processivity along the actin filament. We are interested in how these
two properties are influenced by the bulk concentrations of
nucleotides and the presence of an external force in our model. In
single molecule experiments, all of these parameters can be controlled
and monitored, giving rise to direct measurements of their influence
on the myosin-V walk.

\begin{figure}
  \centering
   \includegraphics[width=130mm]{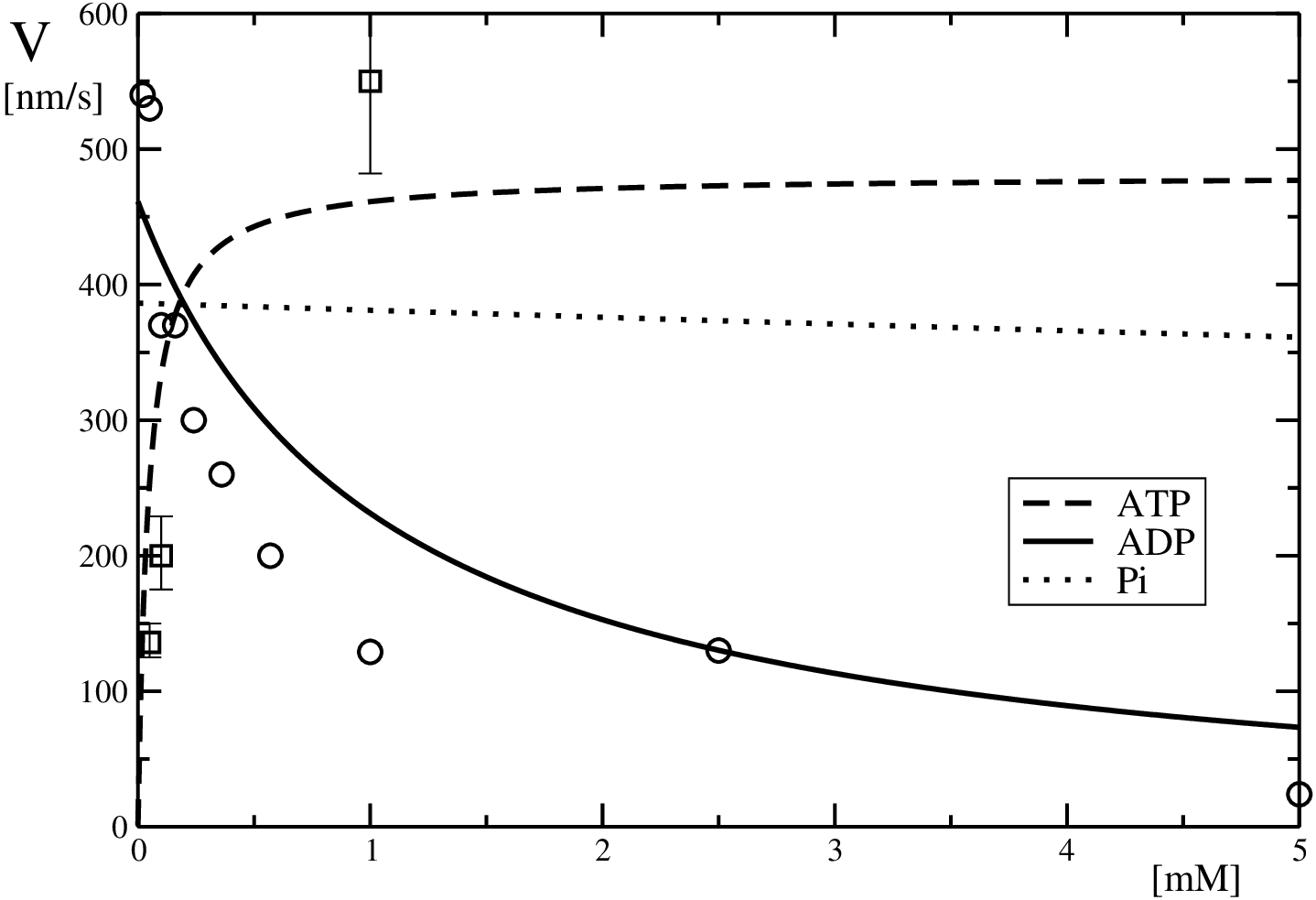}
   \caption{}
        \label{fig:Conc_Velo}
\end{figure}

The velocity of myosin-V has been measured at between 200--500 nm/s
\cite{Mehta2001}, which is consistent with the magnitude of the
velocities predicted by the model. The velocity appears to follow a
Michealis--Menten like form in which it becomes
independent of ATP at high ATP concentration, while linearly dependent
on ATP at low ATP concentrations (Fig.~\ref{fig:Conc_Velo}). There is
similarly found to be a strong dependence of the velocity on the ADP
concentration. ${\rm P_{i}}$ only has a measurable influence on the
velocity (and the run length) at very large excess
concentrations, similarly to what is found experimentally
\cite{Mehta2001,
Baker-Krementsova-Kennedy-Armstrong-Trybus-Warshaw2004pnas}. When
comparing the velocity dependence on concentration of ATP
and ADP directly
with experimental measurements
\cite{Baker-Krementsova-Kennedy-Armstrong-Trybus-Warshaw2004pnas}, we
find reasonable quantitative agreement (Fig.~\ref{fig:Conc_Velo}).

\begin{figure}
  \centering
   \includegraphics[width=130mm]{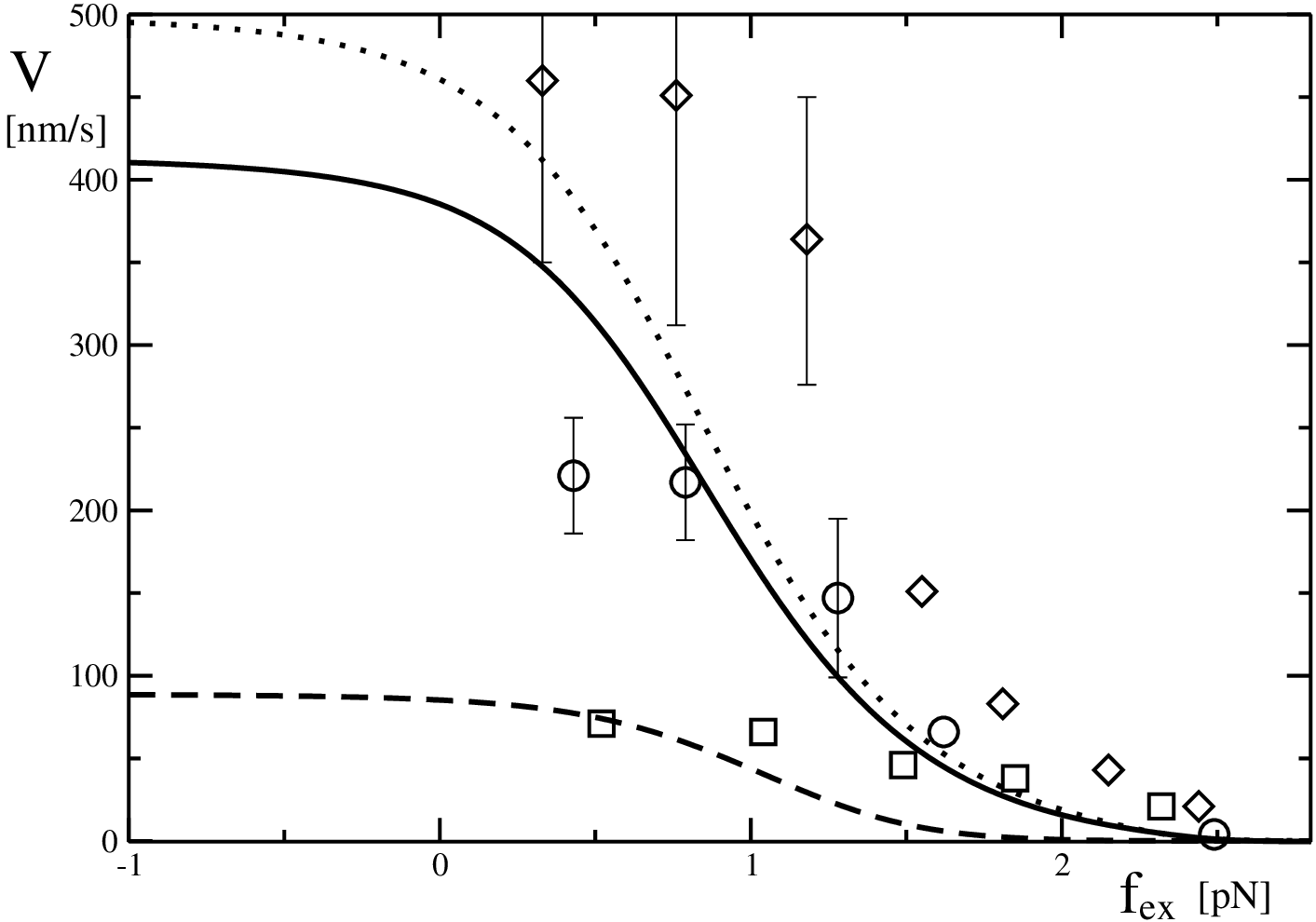}
   \caption{}
        \label{fig:Fex_Velo}
\end{figure}

Our model reproduces the trend of velocity with increasing external force
(see Fig.~\ref{fig:Fex_Velo}) found experimentally in
Ref.~\cite{Uemura-Higuchi-Olivares-DeLaCruz-Ishiwata2004}. 
Note that our model only considers the external force
parallel to the motion of myosin-V along actin, while in an
experiment optical loads are usually applied both along the axis of movement
and perpendicular to it \cite{Fisher-Kim, Kim-Fisher}. This might explain some of the discrepancies observed
at large external forces.
When applying a
negative external force (pulling in the forward direction), the
velocity is not found to increase significantly as has also been found
by \cite{Clemen-Vilfan-Jaud-Zhang-Barmann-Rief2005}. When the pulling 
force in
the direction of the motion exceeds 2 pN, we find that the average 
velocity decreases, since the molecule tends, increasingly, to be pulled off the actin track thereby populating state
7 in the reaction cycle (Fig.~\ref{fig:ModelFig}) and reducing
the forward motion. This reduction in the average velocity at large
negative force was not observed by
\citet{Clemen-Vilfan-Jaud-Zhang-Barmann-Rief2005}. 

\begin{figure}
  \centering
     \includegraphics[width=130mm]{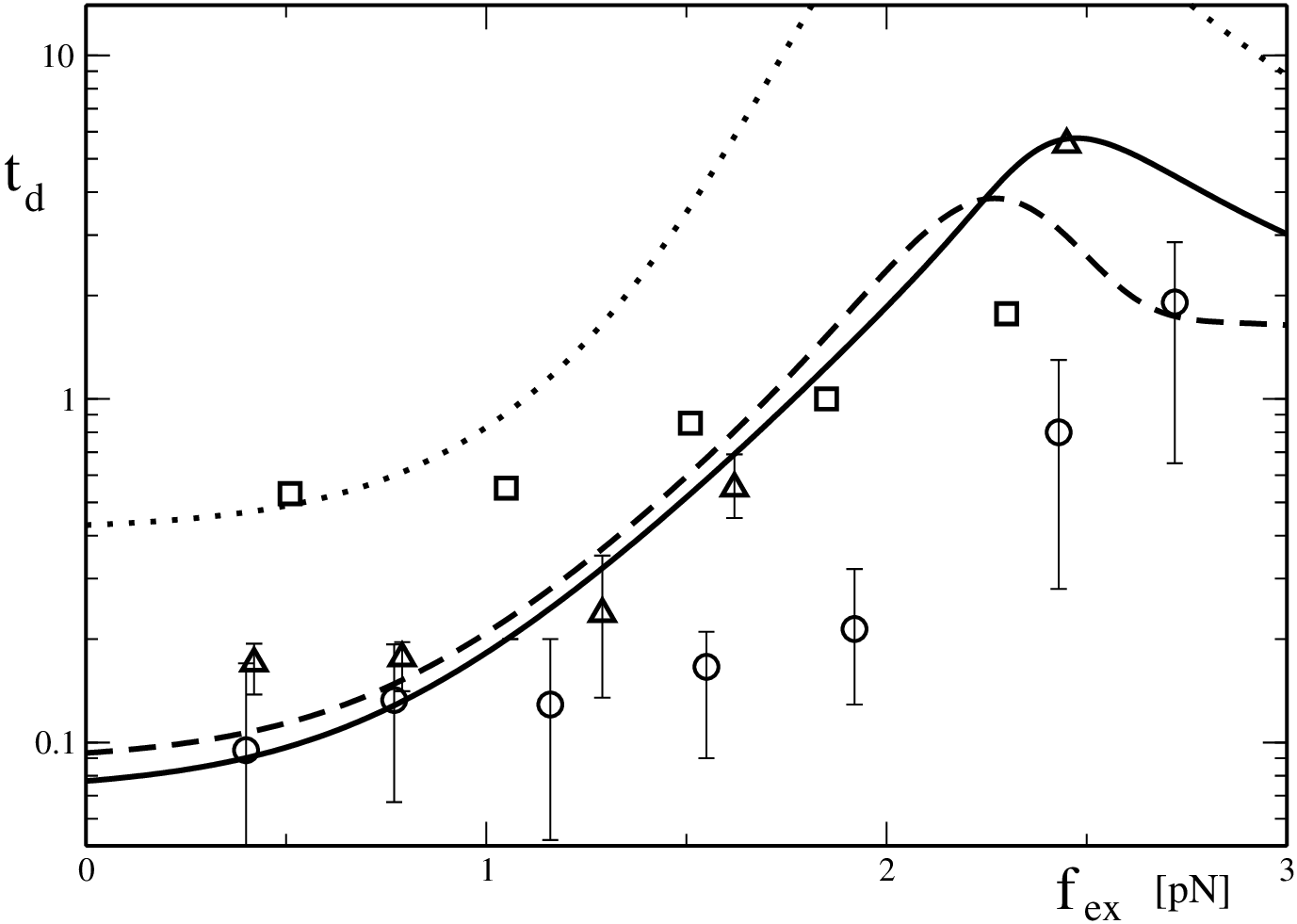}
        \caption{}
     \label{fig:DwellTime}
\end{figure}

Related to the velocity of myosin-V is the dwell time, $t_{\rm d}$,
defined by Eq.~\eqref{eq:dwelltime} in Appendix~\ref{App:Expression}, which is the average time it takes before the molecule takes a \emph{forward}
step \cite{Mehta-Rock-Rief-Spudich-Mooseker-Cheney1999, Kolomeisky-Fisher2003bpj}. Comparing the prediction of the model for the 
dependence dwell time on force with experimental results
\cite{Mehta-Rock-Rief-Spudich-Mooseker-Cheney1999, Uemura-Higuchi-Olivares-DeLaCruz-Ishiwata2004} reveals fair agreement for high ADP and high ATP concentration and the correct trend, though poor quantitative agreement, for high ATP but low ADP concentration (Fig.~\ref{fig:DwellTime}). However, the model is not able
to reproduce the finding \cite{Mehta-Rock-Rief-Spudich-Mooseker-Cheney1999, Uemura-Higuchi-Olivares-DeLaCruz-Ishiwata2004}
that the dwell time becomes independent of force at low ATP concentration.

\begin{figure}
  \centering
     \includegraphics[width=130mm]{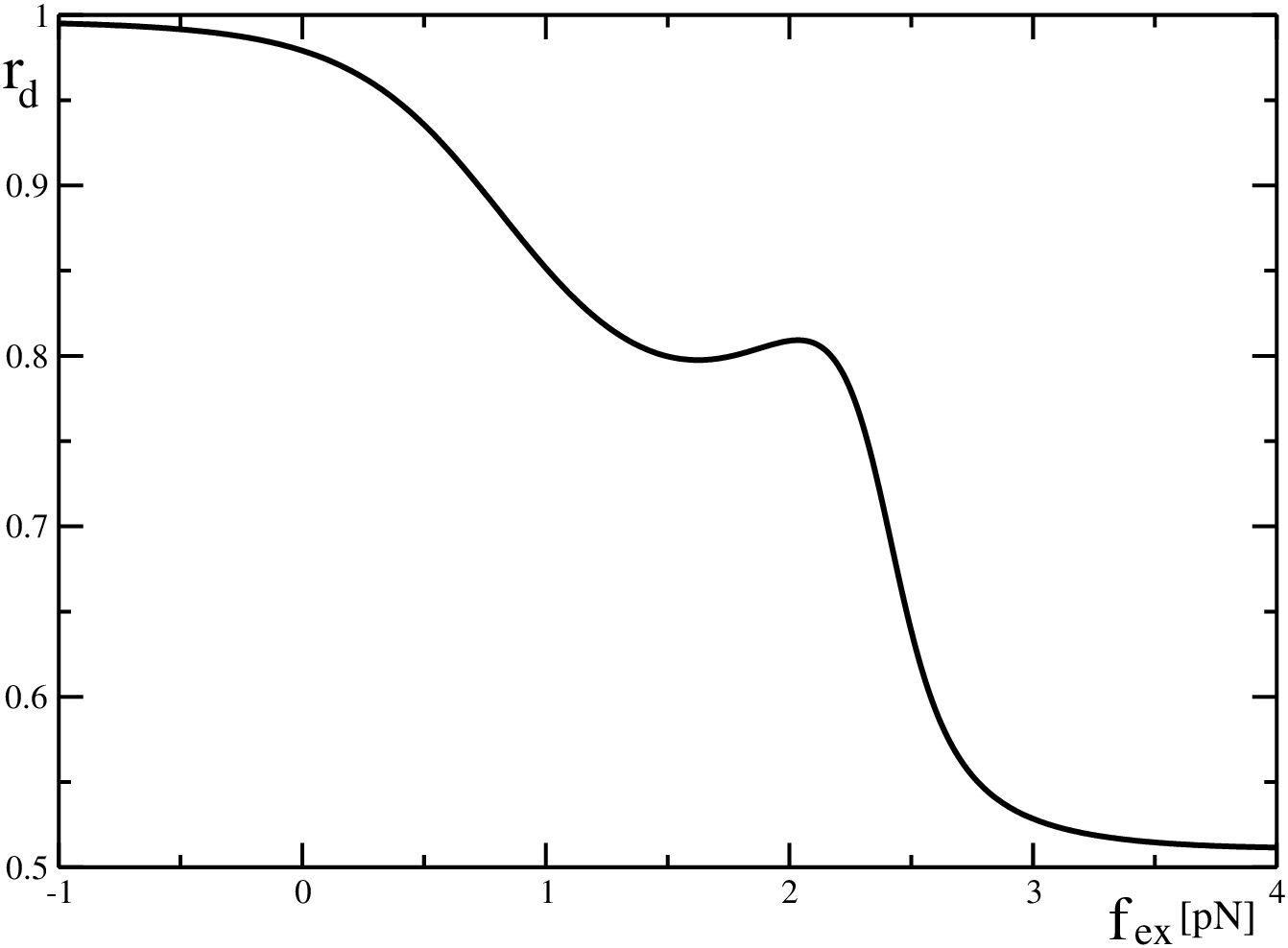}
        \caption{}
      \label{fig:Fex_DutyRatio}
\end{figure}

An important parameter for a processive motor is the duty ratio
\cite{Howard_book}, $r_{\rm d}$, the average proportion of the time the head is
strongly attached to actin.  The duty ratio of myosin-V is found to be
close to 90 \% \cite{Rosenfeld-Sweeney2004}, while our model predicts
even higher duty ratio at low forces
(Fig.~\ref{fig:Fex_DutyRatio}). As expected, the duty ratio is reduced
at higher external force, since the mechanical motion from state 1 to
state 3 is slowed down, and becomes rate-limiting around 1.6 pN at
saturating ATP concentrations, which is consistent with
experimental results of 1.5 pN
\cite{Mehta-Rock-Rief-Spudich-Mooseker-Cheney1999,
Mehta2001}. 

\begin{figure}
  \centering
     \includegraphics[width=130mm]{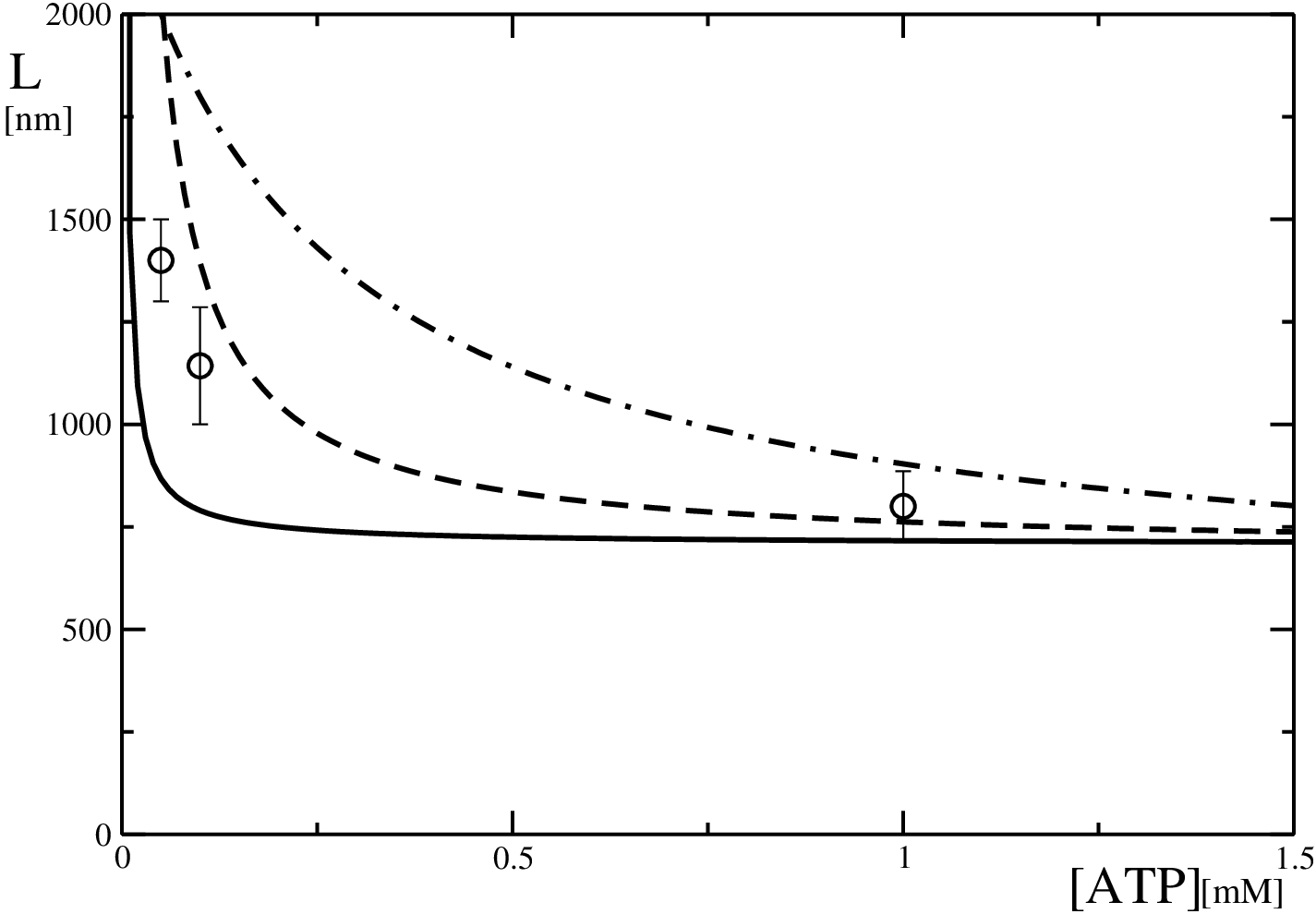}
   \caption{}
      \label{fig:ATP_RunL}
\end{figure}

\citet{Baker-Krementsova-Kennedy-Armstrong-Trybus-Warshaw2004pnas}
have measured run lengths of myosin-V as a function of ADP and ATP
concentrations.  When looking at the dependence of the run length on
the ATP concentration, our model predicts a decrease in run length
when increasing the ATP concentration (Fig.~\ref{fig:ATP_RunL}), which
is consistent with experimental results
\cite{Baker-Krementsova-Kennedy-Armstrong-Trybus-Warshaw2004pnas}.
The run length is found from our model by finding the eigenvalue of
the matrix of rate constants which gives the slowest relaxation time
in the system (see Appendix~\ref{App:Expression}). Since there is one
dominating eigenvalue, we get a single exponential decay in run
length.

We find a nonmonotonic dependency of the run length on the
ADP concentration, as is also found experimentally. (The nonmonotonicity leads to the crossing over of the [ADP] = 1 mM and [ADP] = 1 $\mu$M lines on Fig.~\ref{fig:ATP_RunL} at low ATP concentrations.)
However, while \citet{Baker-Krementsova-Kennedy-Armstrong-Trybus-Warshaw2004pnas}
finds a strong increase in the run length when decreasing the ADP concentration
below [ADP] = 1 mM, our model predicts a decrease in the run length (Fig.~\ref{fig:ATP_RunL}). 
It is not immediately obvious why our model
does not reproduce this experimental finding. It is clear that the
model neglects many possible futile cycles and detachment rates and
maybe even other possible useful reaction
cycles. For instance in \citet{Baker-Krementsova-Kennedy-Armstrong-Trybus-Warshaw2004pnas}
it is suggested that myosin-V needs two useful reaction cycles
to be able to function under variable conditions. Even though our
model only has one main reaction cycle it seems to be able to reproduce many of
the experimental findings under very different conditions, casting some doubt on this earlier assertion. It does not seem at all obvious that it would be of physiological
advantage for myosin-V to reduce its run length under increase of the ADP
concentration. This may suggest that futile cycles not
included in our model play a role. Finally, this puzzle indicates that the ADP dependence of the run length is worthy of further experimental investigation.

\begin{figure}
  \centering
   \includegraphics[width=130mm]{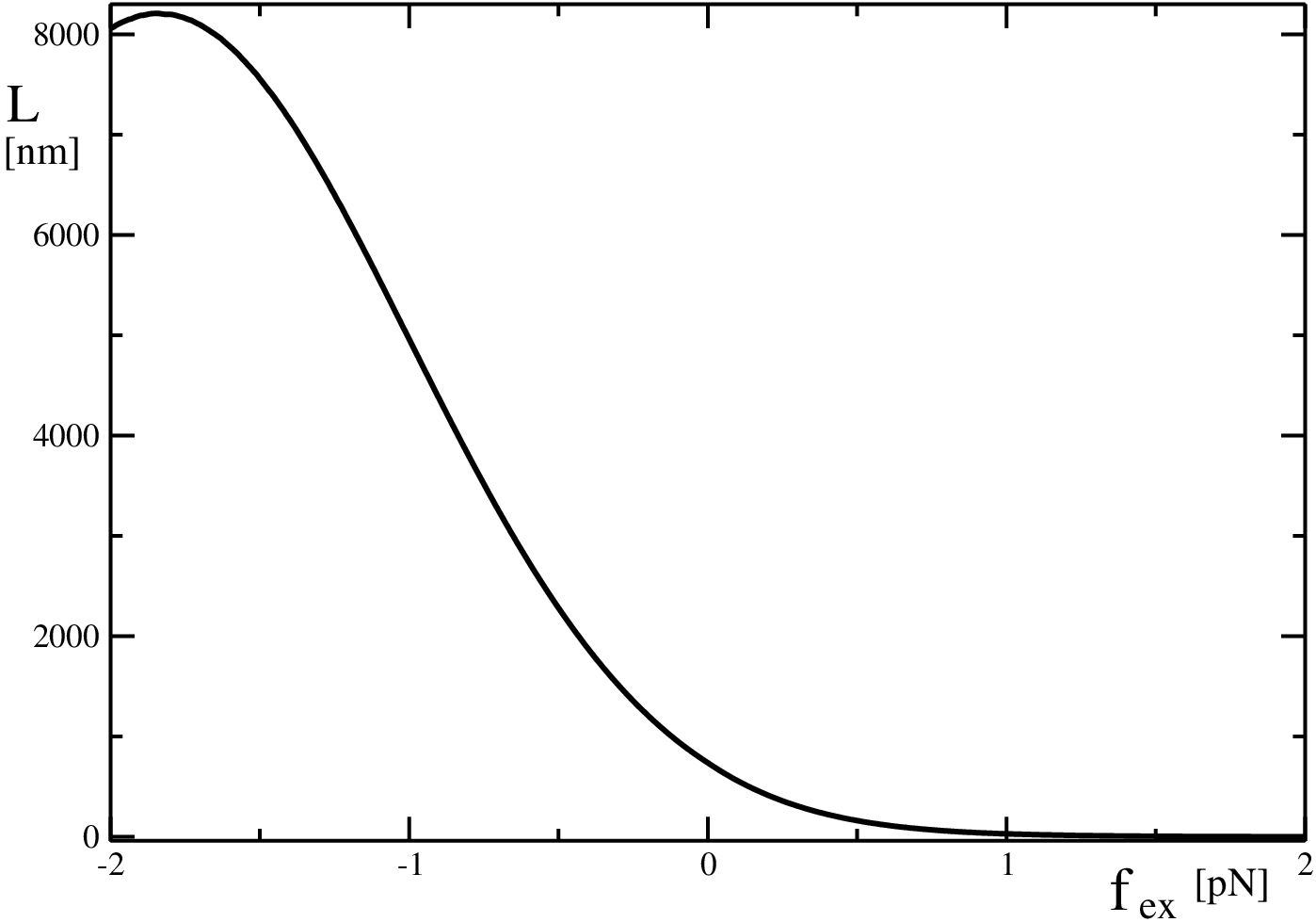}
   \caption{}
        \label{fig:Fex_RunL}
\end{figure}

The run length drops of exponentially when increasing the force as
shown in Fig.~\ref{fig:Fex_RunL}. The run length is also found to be
very influenced by pulling in the \emph{forward} direction (negative
force). This can be understood by considering
Fig.~\ref{fig:Fex_DutyRatio}. The duty ratio increases with an
negative external force since the mechanical movement from state 1 to
state 3 is accelerated. This also makes it less likely that the
molecule will end up in state 7 and detach
(Fig.~\ref{fig:ModelFig}).
At high negative external forces
though, the run length is decreased, since the larger force is pulling
the molecule off the actin track, Eq.~\eqref{eq:detachment_rate}. This
non-monotonic behaviour in the run length as a function of force was
not observed in a recent study
\cite{Clemen-Vilfan-Jaud-Zhang-Barmann-Rief2005} which found 
the run length of myosin-V to be fairly insensitive to
both positive and negative external force over a large range of
values. Note that our model probably exaggerates the increase in attachment
rate of the front head to actin when pulled strongly in the forward direction. For
a strong force the position of
state 2 (Fig.~\ref{fig:Mech}) might change so much that the front head no longer 
is in the target zone slowing down the diffusion to the target site. If a correction for this had been
included in our model we believe that the run length increase for a negative external
force would be reduced. 

\begin{figure}
  \centering
   \includegraphics[width=130mm]{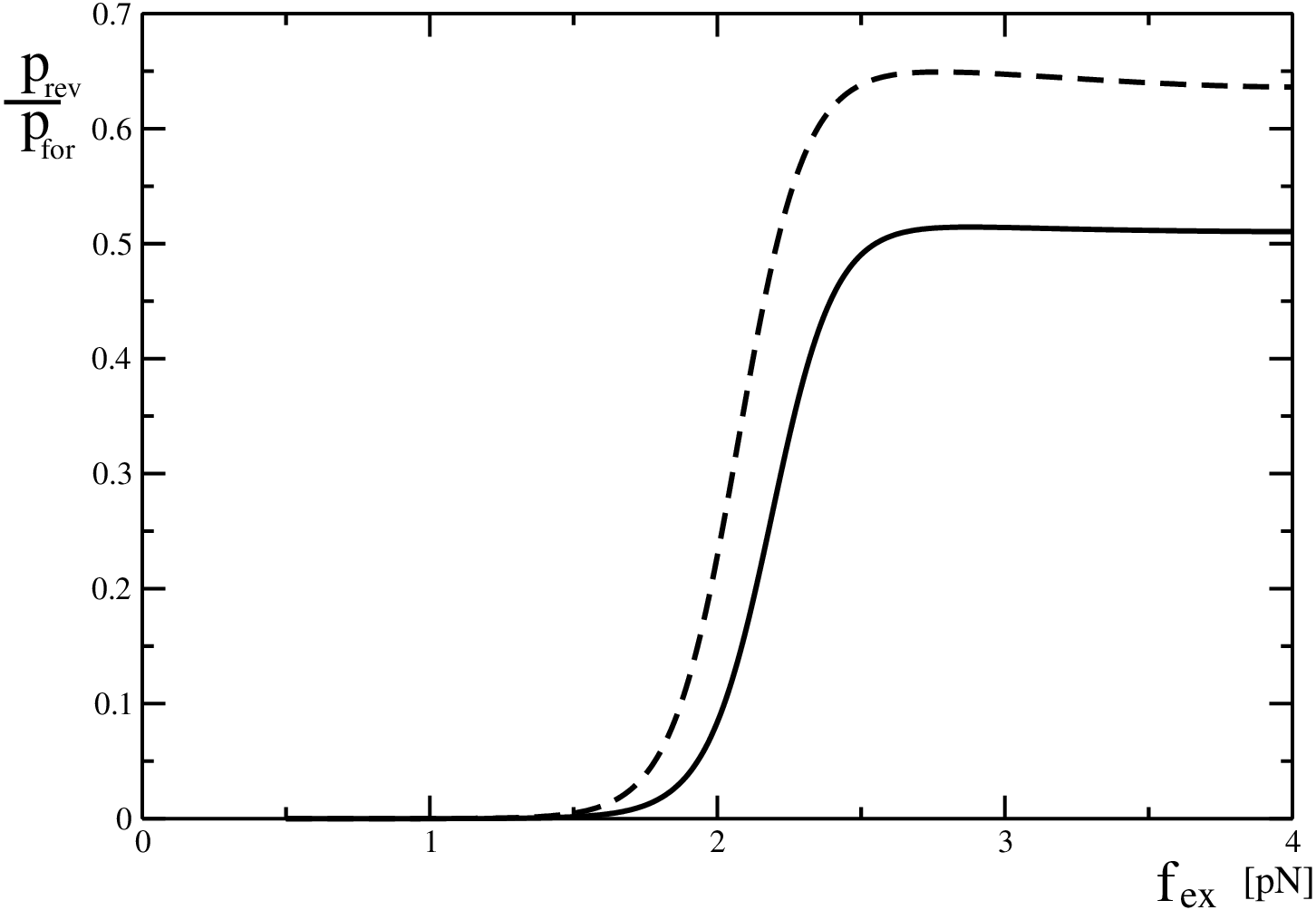}
   \caption{}
        \label{fig:BackFrac}
\end{figure}

Using Eq.~\ref{eq:backwards}, Fig.~\ref{fig:BackFrac} shows the fraction of backwards steps
is found to be low until around 2 pN, consistent with experimental
findings \cite{Rief-Rock-Mehta-Mooseker-Cheney-Spudich2000,
Clemen-Vilfan-Jaud-Zhang-Barmann-Rief2005}. As is shown in 
Fig.~\ref{fig:BackFrac}, the model predicts that
the fraction of backwards steps are larger at low ATP concentrations,
again in agreement with experimental results
\cite{Mehta-Rock-Rief-Spudich-Mooseker-Cheney1999, Mehta2001}.

\begin{figure}
  \centering
     \includegraphics[width=130mm]{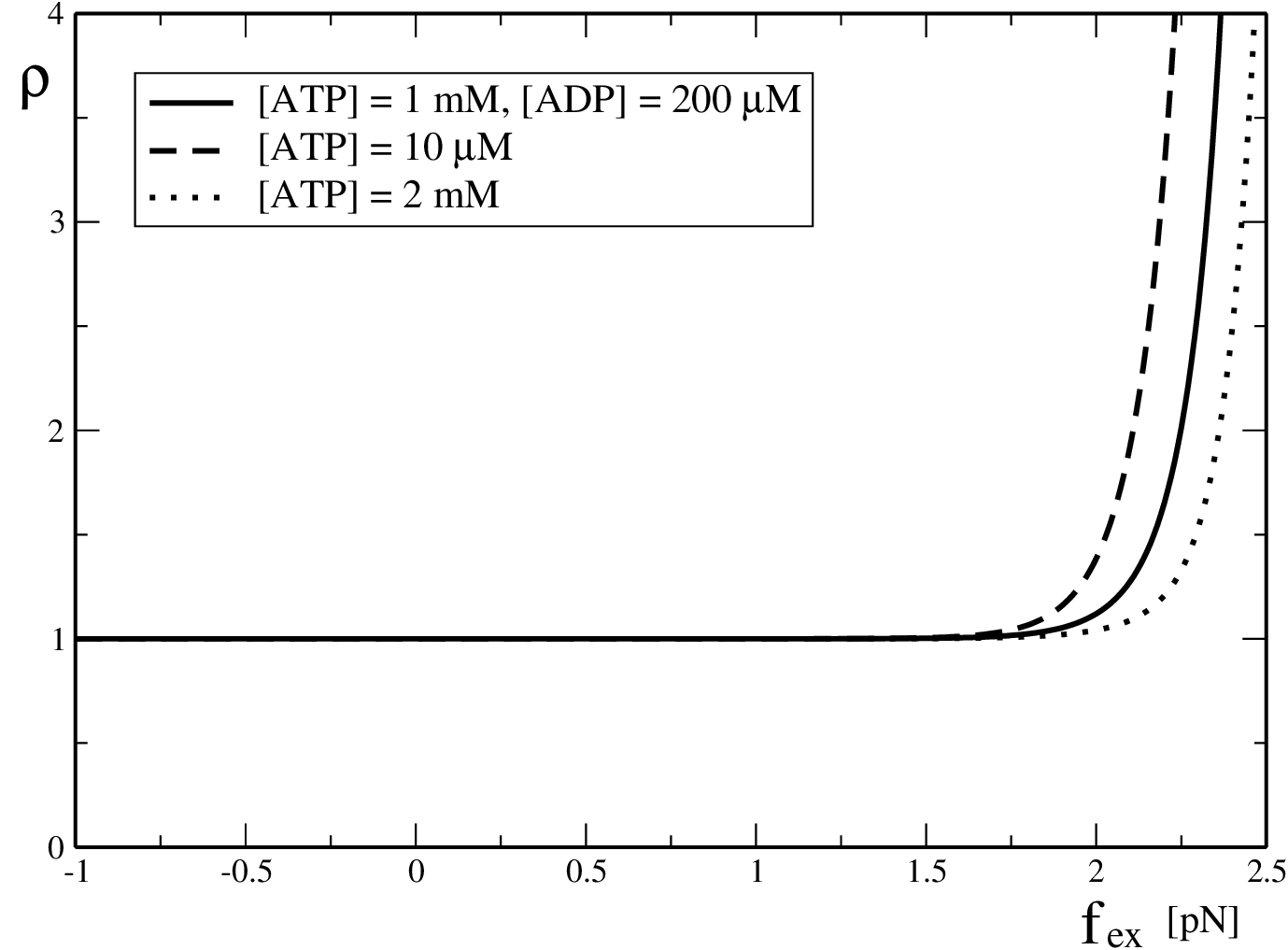}
        \caption{}
     \label{fig:Randomness}
\end{figure}

A quantitative measure of the stochastic deviations from
uniform constant-speed motion (Fig.\ref{fig:Randomness}) is given by the so-called
randomness ratio \cite{Svoboda-Mitra-Block1994, Kolomeisky-Fisher2003bpj}, 
$\rho = 2 D/ V d$, where $D$ is the dispersion
given by Eq.~\ref{eq:disp}. The reciprocal of $\rho$ gives a measure of the number 
of rate limiting steps, and for different ATP concentrations
and external forces, it is found that the model gives only one
rate limiting step. This is somewhat different than earlier
theoretical predictions \cite{Kolomeisky-Fisher2003bpj}, and experimental
measurements of the randomness ratio can be used to differentiate between 
theoretical models. At high forces $\rho$ diverges because of the
vanishing velocity close to stall force.

\begin{figure}
  \centering
     \includegraphics[width=130mm]{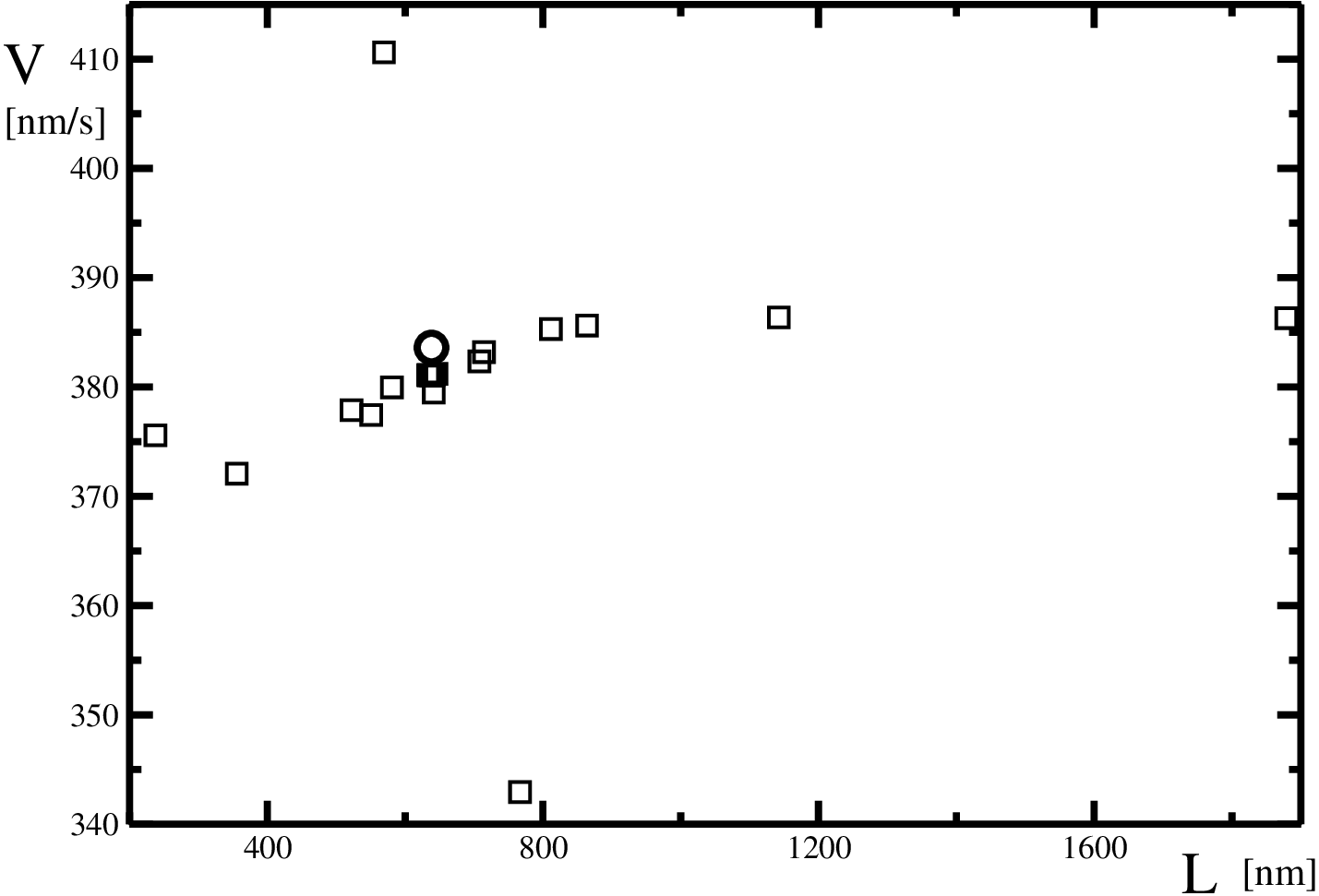}
        \caption{}
       \label{fig:Robustness}
\end{figure}

A biochemical reaction network is expected to be robust to small changes in
the kinetic parameters \cite{Barkai-Leibler1997}. This robustness is
needed to tackle both the natural changes that occur inside a cell
during its lifetime and the fact that cellular biochemical reaction
networks are highly interconnected, so a perturbation in one affects
many others. A simple and effective robustness test is to see how the
motion of the motors is affected by changes of $\pm 5 \%$ in the
different parameters. This also gives information on which parameters
have the largest influence on for instance the velocity and run length 
of the molecule in the model. Fig.~\ref{fig:Robustness} shows that 
$G_4^\ddagger$ and to a lesser degree $E_{\rm strain}$ have the largest influence on the
run length, while $\Delta G_4$ is the only parameter which gives significant
changes in the velocity when perturbed. The importance of 
$G_4^\ddagger$ and $\Delta G_4$ in controlling velocity and run length is 
to be expected, since these are the parameters determining the rate constants
$u_4$ and $w_4$ for ADP release from and recapture by the rear head between
states 4 and 5, which we have found to be the rate-limiting step.

Since the temperature enters explicitly in our equations, via the thermal energy scale $k_{\rm B} T$, it is
natural to look at how the temperature influences the run length and
the velocity.  Some caution should be noted though, since it is found
that some of the free energy terms themselves are known to be directly dependent on temperature
\cite{Rosenfeld-Sweeney2004}. Fig.~\ref{fig:Temp} is obtained under
the assumption that none of the energy terms depend strongly on
temperature.

% We can also see what happens when the bulk viscosity increases, which
% will give indication of what happens when doing the experiment in vivo
% rather than in vitro.

% By varying the internal strain $E_{\rm strain}$ and the stepping
% length $d$, one can make predictions on how changing the number of IQ
% motifs in the neck region should influence the walk.

% --- --- --- --- section* : Discussion   --- --- --- --- --- --- --- --- --- %
\section*{Discussion}

Many reasonable and well justified models of the myosin-V's walk exist in the
literature, but the majority of the models are of a qualitative nature 
and often introduce unnecessary (and uncontrolled) approximations 
when employed to obtain quantitative predictions. Of quantitative models in
the spirit of this work, one interesting early study for myosin-V is due to
\citet{Kolomeisky-Fisher2003bpj} who considered a two state model which took some account of step-size
variations. It was found that there is a substep in the walk
of myosin-V creating 
another possible reaction pathway, as also suggested in some experiments 
\cite{Moore-Krementsova-Trybus-Warshaw2001,Uemura-Higuchi-Olivares-DeLaCruz-Ishiwata2004}. One shortcoming in our model, 
is that the model does not allow for the
possibility that the geometry or mechanics of the walk can change in
different regimes, for instance at very different bulk concentrations
or external forces.  It is still an open question whether motor
proteins change behaviour in a dramatic way in different regimes, but
some evidence suggests that stepping length is influenced by force
\cite{Veigel-Wang-Bartoo-Sellers-Molloy2002,
Uemura-Higuchi-Olivares-DeLaCruz-Ishiwata2004,
Clemen-Vilfan-Jaud-Zhang-Barmann-Rief2005}. Some recent papers \cite{Vilfan2005, Lan-Sum2005} 
have tried to explicitly calculate the different strain energies
for the different step lengths, which would be helpful to clarify
the detailed mechanism of the motion of myosin-V. 
One problem in such calculations is that
several angles and rigidities are not well known so that the number of free 
parameters becomes very large.

There is necessarily some arbitrariness as to how many and which
distinct states are included in a model, since the concept is strictly
only a useful {\em approximation} to the complex, fluctuating motion
of the protein. Several sub-states (and thereby sub-steps in the main
reaction cycle) which have been identified experimentally or on
thermodynamic grounds \cite{Geeves-Goody-Gutfreund1984,
Houdusse-Sweeney2001, DeLaCruz-Wells-Sweeney-Ostap2000} are not
included in this model. This means that the sub-reaction rates between
some of the states have been combined into one effective rate.  Where possible we
have chosen not to excessively ``coarse-grain" the state-space, e.g. by classifying many states together into fewer, more broadly defined states. Philosophically this seems wise if one is not certain {\it a priori} that such coarse graining of state space won't reduce the models precision or predictive power. We also feel that it is of value that the model takes into account all
the steps where the protein reacts with smaller bulk molecules, since
bulk concentration is something that can be controlled
experimentally. Similarly, we wanted to separate out the
different steps that are expected to couple to the external force. As
a result, the proposed model can make direct predictions as to how the
external force produced by an optical tweezer, for instance, should
change the behaviour of the myosin-V walk. It is anyway important to
remember that the complexity of a model such as ours is not really a function of the number of
states but rather the number of parameters, which remains small, the
interpretation of the parameters, which remains physically clear, and the reaction network topology.

The elastic strain in the molecule plays several roles in our model;
most importantly, the strain in the molecule synchronises the chemical
reactions of the two myosin heads. This is vital for making myosin-V
an effective processive motor
\cite{Veigel-Wang-Bartoo-Sellers-Molloy2002}. The synchronisation is
caused by the slowing down of the ADP release from the lead head
compared to the rear head in state 4, since the internal strain makes
it less likely for the lead head to have the optimal angle relative to
the actin filament necessary for ADP release.  A related benefit of
this is that the slowing down of ADP release from the lead head
minimises the impact of the futile cycle. Also, the strain biases the
detached head to stay in the ``target zone'' (state 2,
Figs.~\ref{fig:ModelFig} and \ref{fig:Cycle}) when doing the (11 nm) biased diffusional
search for the next attachment site on the actin filament.  This bias
both increases the velocity of the motor, since less time is used in
the diffusional search for the next target site on the actin, and
increases the processivity by decreasing the time spent with only one
head strongly attached to the actin.

The model does not contain any direct dissipation of free energy
$\Delta G_{\rm diss}$.  Such dissipation by friction and heat loss
caused by the non-equilibrium motion of the protein would reduce the
total useful work that the molecular motor could do by reducing the
effective free energy of the hydrolysis reaction $\Delta G_{\rm eff} =
\Delta G_{\rm hyd} - \Delta G_{\rm diss}$. We have not tried to
quantify the size of $\Delta G_{\rm diss}$ but assumed that it is
relatively small. It would of course be possible to include $\Delta
G_{\rm diss}$ as an undetermined parameter even though we have chosen
not to do this here. The model does take into account the energy
dissipated by the futile hydrolysis of ATP by the futile cycle
(Fig.~\ref{fig:ModelFig}) which does no work and creates no net
movement, but since $\alpha E_{\rm strain}$ is estimated to be
relative large, the release rate of ADP from the front head in state 4
(Fig.~\ref{fig:ModelFig}) is reduced two orders of magnitude
compared to the release of the rear head. This indicates that myosin-V
is a tightly coupled motor under all conditions and the futile cycle
will not dissipate significant energy.

The starting point for our model was the qualitative model proposed by
\citet{Rief-Rock-Mehta-Mooseker-Cheney-Spudich2000} and the
observation that myosin-V has a so-called powerstroke movement and a
diffusional search. There are several alternative mechanisms suggested
in the literature, although they typically have a lot in common since
there is general agreement on the single head reaction mechanism. One
influential alternative model has been proposed by
\citet{DeLaCruz-Ostap-Sweeney2001} where the strong attachment of the
front head to actin is triggered by the release of the rear head.  It
is also possible to envision there being several \emph{parallel}
reaction paths followed by the motor, where all these paths contribute
significantly to the forward motion. It would be natural to assume
that different paths dominate in different regimes, which would then
provide a strategy for the motor to function well under varying
conditions. One interesting example
\cite{Baker-Krementsova-Kennedy-Armstrong-Trybus-Warshaw2004pnas}
of such a parallel reaction path mechanism, involved combining the Rief {\it et al.} and the
De La Cruz {\it et al.} models into a version with two-paths. Taking into account the
number of proposed models in the literature and then the possibility
of different combinations of the various models, the number of
permutations of possible models is clearly very large. We argue that a more 
quantitative analysis of the competing models would often be useful, 
e.g.~using a computational scheme similar to that presented here.

We have presented a moderately detailed model of the walk of myosin-V,
and compared the accuracy of the model with experimental measurements
for velocities, run lengths, and dwell times at different
nucleotide concentrations and external force. Predictions are also
made for as yet unmeasured quantities, such as the internal strain in 
the molecule and the randomness ratio. The model also clearly shows how the
internal strain can be used by myosin-V to coordinate its forward
motion. A clear advantage of our model is that the physical significance of parameters
in the model is transparent. This transparency makes the model a
useful reference for comparison to future experiments and aids in the
identification of elements of the model which are accurate and
elements that need refinement.  As our work and other recent quantitative
studies have demonstrated, it is possible to give a clear analysis of
very detailed models. Even when experimental data are individually not conclusive,
the large number of existing experimental measurements available should help to
differentiate the models when they are analysed in such detail. In turn
this will be useful in clarifying the underlying
mechanism of the myosin-V walk, which is still not precisely understood.

We remark that ours is the first attempt to explore carefully whether a full reaction cycle, with a nearly complete list of relevant parameters, is or is not able to reproduce the trends observed over a variety of experimental studies. There are several useful outcomes from our study. With the exception of the dependence of run length on ADP concentration our model does appear to be broadly consistent with the experimental observations. We are still unable to explain even the qualitative nature of the run length dependence on ADP concentration. This is a puzzle, which might signify a flaw in the accepted models and which we flag for future experimental attention.  In addition, we find that phosphate release must be coupled to a release of energy over much of the transition between states 3 and 4 if the model is to produce results that are consistent with known rates of Pi release. 
\vskip 15pt
\noindent
{\Large\bf Acknowledgments}
\vskip 10pt
\noindent
The authors would like to acknowledge the enormous assistance afforded by one particularly scrupulous and knowledgeable referee. We have also benefitted from discussions with Prof Kolomeisky (Rice), Prof Sun (Hopkins) and Prof Molloy (Mill Hill, London) who was also inspirational in motivating this work. We acknowledge the support of EPSRC through grant GR/S24671/01.

% *** * * * * * * * * * * * * *  APPENDICES * * * * * * * * * * * * * * * *** %
    \appendix

% * * * * Appendix :  Velocity Expression   * * * * * * * ** * * * * * * * * %
\section{Analytical Expressions}
\label{App:Expression}

These expressions were used to compare with experimental data.

The master equations for the probabilities, $P_j$, of finding the
molecule in state $j$, can be deduced from Fig.~\ref{fig:Cycle} to be
\begin{eqnarray}
  \dot{P_{1}} &=& u_{5}\, P_{5} + w_{2} \, P_{2} - (u_1 + w_1) \, P_{1}
               \,,  \\
  \dot{P_{2}} &=& u_{1} P_{1} + w_{3} P_{3} - (u_2 + w_2 + u_{2,7} + w_{2,6}) P_2 + u_{6,2} P_6 + w_{7,2} P_7
               \,,  \\
  \dot{P_{3}} &=& u_{2}\, P_{2} + w_{4} \, P_{4} - (u_3 + w_3) \, P_{3}
               \,,  \\
  \dot{P_{4}} &=& u_{3}\, P_{3} + w_{5}\, P_{5} - (u_4 + w_4 + u_{4,6})\, P_4 + w_{6,4}\, P_6
               \,, \\
  \dot{P_{5}} &=& u_{4}\, P_{4} + w_{1} \, P_{1} - (u_5 + w_5) \, P_{5}
               \,,  \\
  \dot{P_{6}} &=& u_{4,6} \, P_4 + w_{2,6} \, P_2 - (u_{6,2} + w_{6,4})\, P_6
               \,, \\
  \dot{P_{7}} &=& u_{2,7}\, P_{2} - (w_{7,2} + u_7)\, P_{7} 
                \,. \label{eq:P_7} 
\end{eqnarray}
In matrix notation we have
\begin{equation} \label{eq:P_mat}
  \dot {\mathbf P} = {\mathbb M}\,{\mathbf P} \,, 
\end{equation}
where ${\mathbb M}$ is the 7x7 reaction rate matrix and the $j$th
component of the vector ${\mathbf P}$ is $P_j$.

The probability of finding the molecule in one of states $1$ to $7$ is
not conserved under equation (\ref{eq:P_mat}) since the motor drops
off the actin track at a rate $u_7$ from state $7$ (equation
(\ref{eq:P_7})). In order to solve equation (\ref{eq:P_mat}) it is
helpful to renormalise, eliminating drop-off, to get a
probability-conserving equation. This is done by writing
\begin{equation} \label{renorm}
  P_j = \frac{1}{\varphi_j} \, {\rm e}^{-\lambda t}\,\tilde{ P_j} \,
\end{equation}
and choosing the constants $\varphi_j$ such that the renormalised
probabilities $\tilde{P}_j$ satisfy a conservative set of
equations. It is possible to show that this can be done (see
Ref.~\cite{Kolomeisky-Fisher2000pA}) if
\begin{equation} \label{eig}
  {\mathbb M}^{T}\,{\boldsymbol \varphi} = -\lambda \,{\boldsymbol \varphi} \,, 
\end{equation}
and that then the vector $\tilde{{\mathbf P}}$ of 
renormalised probabilities satisfies the equation 
\begin{equation} \label{eq:P_matre}
  \dot{\tilde{{\mathbf P}}} = \tilde{{\mathbb M}}\,\tilde{{\mathbf P}} \,, 
\end{equation}
where $\tilde{{\mathbb M}}$ is a renormalised reaction-rate matrix 
with 
$\tilde{u}_{7} = 0$ and renormalised rate constants
\begin{equation}
  \tilde{u}_j = u_j \frac{\varphi_{j+1}}{\varphi_{j}} \qquad {\rm and}\qquad 
  \tilde{w}_j = w_j \frac{\varphi_{j-1}}{\varphi_j} \,,
\end{equation}
for $j=1\ldots 5$, where the index is periodic with period $5$, and 
\begin{equation}
  \tilde{u}_{i,j} = u_{i,j} \frac{\varphi_{j}}{\varphi_{i}} \qquad {\rm and}\qquad 
  \tilde{w}_{i,j} = w_{i,j} \frac{\varphi_{j}}{\varphi_i} \,,
\end{equation}
for the rate constants not on the main reaction cycle.

The slowest, dominant eigenvalue, $-\lambda_0$, is negative for
reasonable rate constants, while the faster eigenvalues can be complex
giving rise to fast oscillations in the reaction network.

At long times, $\tilde{{\mathbf P}}$ tends to the steady-state
solution of equation (\ref{eq:P_matre})
\cite{Kolomeisky-Fisher2000pA}, found by solving $\tilde{{\mathbb
M}}\,\tilde{{\mathbf P}} = 0$ analytically.  The relaxation times of
oscillations are found to be fast compared to the stepping time and to
$1/\lambda_0$, so it is clear from equation (\ref{renorm}) that the
solution for $P_j$ at long times will be dominated by the slowest
eigenvalue, $-\lambda_0$. We therefore choose ${\boldsymbol \varphi}$
to be the corresponding eigenvector of the transposed matrix ${\mathbb
M}^{T}$. Since there are seven states in our model, this eigenvalue
problem is best solved numerically.

In the following equations we assume all the rate constants are
renormalised and drop the tilde. We look for steady-state solutions
$\dot{P_j}=0$. (For another approach to achieve analytical expressions
for a fairly similar reaction network, see \citet{Kolomeisky2001}). 
The net flux between two neighbouring states is given
by $J = P_j u_j - P_{j+1} w_{j+1}$, and we seek to express the
steady-state probabilities $P_j$ in terms of the futile flux $J_{\rm
fut} = P_{6} u_{6,2} - P_{2} w_{2,6}$ and the main reaction flux
$J_{\rm hyd} = V/d = P_{1} u_{1} - P_{2} w_{2}$. It can be
shown that the steady-state solution is given by
\begin{equation}
  P_j = \frac{\Gamma}{\Gamma - 1}\biggl[ 
    r_j \frac{V}{d}  + s_j J_{\rm fut}
    \biggr] \,,
\end{equation}
for $j=1,...,5$, $P_7= (u_{2,7}/w_{7,2}) P_2$, and
\begin{equation}
  P_6 = \frac{\Gamma_2}{\Gamma_2 - 1}\biggl[ 
    r_1^{\, \prime} J_{\rm fut} + s_1^{\, \prime} \frac{V}{d}
    \biggr] \,,
\end{equation}
where $\Gamma$ is given by Eq.~\eqref{eq:Gamma} and
\begin{equation}
\Gamma_2 = \prod_{j=1}^{4}  \frac{u_j^{\, \prime}}{w_j^{\, \prime}} 
= \frac{u_{6,2}\, u_2\, u_3\, u_{4,6}}{w_{6,4}\, w_{2,6}\, w_3\, w_4} = {\rm e}^{\Delta G_{\rm hyd}/ k_{\rm B} T}\,.
\end{equation}
Here we have introduced the auxiliary functions
\begin{eqnarray}
  r_j &=& \frac{1}{u_j}\, \biggl( 1 + \sum_{k=1}^{4} \prod_{i=1}^{k}
      \frac{w_{j+i}}{u_{j+i}} \biggr) \,, 
                   \label{eq:r_j} \\
  s_j &=& \frac{1}{u_j}\, 
      \biggl( (\delta_{j,2} + \delta_{j,3}) 
      + \sum_{k=1}^{4} (\delta_{j+k,2} + \delta_{j+k,3})
      \prod_{i=1}^{k} \frac{w_{j+i}}{u_{j+i}} 
       \biggr)   \,,
                    \label{eq:s_j}
\end{eqnarray}
for $j=1,...,5$, where $\delta_{j,i}$ is the Kronecker delta function
(with $\delta_{j,i} \equiv \delta_{j+5,i}$) and where the indices on
the rate constants are periodic with period 5. We have similar
auxiliary functions relating to the futile cycle, namely
\begin{eqnarray}
  r_j^{\, \prime} &=& \frac{1}{u_j^{\, \prime}}\, \biggl( 1 +
      \sum_{k=1}^{3} \prod_{i=1}^{k} \frac{w_{j+i}^{\,
      \prime}}{u_{j+i}^{\, \prime}} \biggr) \,, 
               \label{eq:r_j^'} \\
  s_j^{\, \prime} &=& \frac{1}{u_j^{\, \prime}}\, \biggl(
      (\delta_{j,2} + \delta_{j,3}) + \sum_{k=1}^{3} (\delta_{j+k,2} +
      \delta_{j+k,3}) \prod_{i=1}^{k} \frac{w_{j+i}^{\,
      \prime}}{u_{j+i}^{\, \prime}} \biggr) \,, 
                \label{eq:s_j^'}
\end{eqnarray}
for $j=1,...,4$, where $w_{j}^{\, \prime}=w_{j}$ and $u_{j}^{\,
\prime}=u_{j}$ except for $w_{1}^{\, \prime} = w_{6,4}$, $w_{2}^{\,
\prime}=w_{2,6}$, $u_{1}^{\, \prime}=u_{6,2}$, and $u_{4}^{\,
\prime}=u_{4,6}$. We now have a period of four, $u_j^{\, \prime}\equiv
u_{j+4}^{\, \prime}$, $w_j^{\, \prime}\equiv w_{j+4}^{\, \prime}$, and
$\delta_{j,i}\equiv \delta_{j+4,i}$, since there are four states that
participate in the futile cycle (Fig.~\ref{fig:ModelFig}).

Using the conservation of probability $\sum P_j = 1$ (since drop-off
has been eliminated through renormalisation) and a second independent
expression for $P_4$,
\begin{equation}
  P_4 = \frac{\Gamma_2}{\Gamma_2 - 1}\biggl[ 
    r_4^{\, \prime} J_{\rm fut} + s_4^{\, \prime} \frac{V}{d}
    \biggr] \,,
\end{equation}
one can derive the analytical expressions for the velocity, $V$, and the futile flux, $J_{\rm fut}$:
\begin{eqnarray}
  V &=&  B\,d
    \biggl(
      \frac{r_4^{\, \prime}}{\Gamma(\Gamma_2 - 1)} - \frac{s_4}{\Gamma_2(\Gamma - 1)} 
        \biggr)\,, \\
  J_{\rm fut} &=&  B\,
    \biggl(
      \frac{r_4}{\Gamma_2(\Gamma - 1)} - \frac{s_4^{\, \prime}}{\Gamma(\Gamma_2 - 1)} 
        \biggr)\,, \label{eq:futflux}
\end{eqnarray}

where
\begin{align}
  \frac{1}{B} = \Gamma\,\Gamma_2\biggl[\, 
    &\biggl(\frac{\sum_{j=1}^{5} \tilde{r}_j}{\Gamma_2(\Gamma - 1)} 
    + \frac{s_1^{\, \prime}}{\Gamma(\Gamma_2 - 1)}\biggr)
  \biggl(\frac{r_4^{\, \prime}}{\Gamma(\Gamma_2 - 1)} - \frac{s_4}{\Gamma_2(\Gamma - 1)} \biggr)
    \nonumber \\
  &+ \biggl(\frac{\sum_{j=1}^{5} \tilde{s}_j}{\Gamma_2(\Gamma - 1)} + \frac{r_1^{\, \prime}}{\Gamma(\Gamma_2 - 1)}\biggr)
  \biggl(\frac{r_4}{\Gamma_2(\Gamma - 1)} - \frac{s_4^{\, \prime}}{\Gamma(\Gamma_2 - 1)} \biggr) \biggl] \,,
\end{align}
with $\tilde{r}_j=r_j$ and $\tilde{s}_j=s_j$ for $j=1,3,4,5$, while
$\tilde{r}_2=r_2(1+u_{2,7}/w_{7,2})$ and
$\tilde{s}_2=s_2(1+u_{2,7}/w_{7,2})$.
% If the two reaction paths are completely independent, i.e.~$s_j = 0$, the expressions reduce to expressions derived
% by Kolomeisky and Fisher \cite{}.

With the probability of finding the molecule in each state now
determined, we can use previously established analytical
expressions. The fraction of backward steps is
\cite{Kolomeisky-Fisher2003bpj, vanKampen_book}
\begin{equation}
  \pi_{-} = \frac{\hat{w}_0 w_1}{\hat{u}_0 u_1 + \hat{w}_{0} w_{1}} \,,
	\label{eq:backwards}
\end{equation}
where $\hat{u}_0 = u_5 \,\bigl(P_5/\sum_{j=2}^{7} P_j\bigr)$ and
$\hat{w}_0 = w_2 \, \bigl(P_2/\sum_{j=2}^{7} P_j\bigr)$.

The mean forward-step dwell time is given by
 \cite{Kolomeisky-Fisher2003bpj, vanKampen_book}
\begin{equation}
  t_{\rm d} = \frac{\hat{u}_0 + u_1 + \hat{w}_0 + w_1}{\hat{u}_0 u_1 + \hat{w}_0 w_1} \,,
    \label{eq:dwelltime}
\end{equation}
and the dispersion is given by \cite{Fisher-Kolomeisky1999pA,
Derrida1983}
\begin{equation}
  D =  \frac{1}{2} \biggl[\frac{\hat{u}_0 u_1}{\hat{w}_0 w_1} + 1 - 2 \left (\frac{\hat{u}_0 u_1}{\hat{w}_0 w_1} - 1 \right )
    \frac{\hat{w}_0 w_1}{(\hat{u}_0 + u_1 + \hat{w}_0 + w_1)^2}\biggr] \frac{\hat{w}_0 w_1}{(\hat{u}_0 + u_1 + \hat{w}_0 + w_1)} d^2   \,.
    \label{eq:disp}
\end{equation}

Since the probability of the myosin remaining attached to the actin filament decays exponentially with the dominant eigenvalue $-\lambda_0$, the typical duration of a run is $1/\lambda_0$ and hence the run length $L$ is given by
\begin{equation}
L= \frac{V}{\lambda_0}.
\end{equation}
%

% * * * * Appendix :  Cost Function         * * * * * * * ** * * * * * * * * %

\section{Cost Function}
  \label{App:CostFun}

The cost function contains 17 terms. Except where other
concentrations are explicitly mentioned, the nucleotide concentrations
are given by [ATP] = 1 mM and [ADP] =[${\rm P_i}$] = 0.1 $\mu$M. Likewise, the external force is assumed to be zero, except if
stated otherwise. 

The first term in the cost function is constructed from the model
velocity, $V$, compared to the velocity $V^E$ estimated from experimental data
in the literature \cite{Mehta-Rock-Rief-Spudich-Mooseker-Cheney1999,
DeLaCruz-Wells-Rosenfeld-Ostap-Sweeney1999,
DeLaCruz-Sweeney-Ostap2000,
Rief-Rock-Mehta-Mooseker-Cheney-Spudich2000,
Uemura-Higuchi-Olivares-DeLaCruz-Ishiwata2004,
Baker-Krementsova-Kennedy-Armstrong-Trybus-Warshaw2004pnas}. The
mean squared uncertainty in the measured velocities, $\sigma_{V^E}^2$, are 
in general found to be around 10\%. 
\begin{equation}
  \Delta^{(1)} = \frac{\bigl[V - V^E\bigr]^2}{\sigma_{V^E}^2} = \frac{\bigl[V - 540 \,{\rm nm/s} \bigr]^2}{ (54 \,{\rm nm/s})^2 } \,.
\end{equation}	              
When the ADP concentration increases the velocity $V^E$ is found to
decrease \cite{Rief-Rock-Mehta-Mooseker-Cheney-Spudich2000,
DeLaCruz-Wells-Rosenfeld-Ostap-Sweeney1999,
DeLaCruz-Sweeney-Ostap2000,
Wang-Chen-Arcucci-Harvey-Bowers-Xu-Hammer-Sellers2000,
Baker-Krementsova-Kennedy-Armstrong-Trybus-Warshaw2004pnas}. We choose
a second and third ADP concentration, $[{\rm ADP}]^{(2)}$ = 200 $\mu$M
and $[{\rm ADP}]^{(3)}$ = 2.5 mM, and construct additional terms
in the cost function at these conditions:
\begin{equation}
  \Delta^{(2)} = \frac{\bigl[V([{\rm ADP}]^{(2)}) - V^E([{\rm ADP}]^{(2)})\bigr]^2}{\sigma_{V^E([{\rm ADP}]^{(2)})}^2} 
    = \frac{\bigl[V([{\rm ADP}]^{(2)}) - 320 \,{\rm nm/s} \bigr]^2}{ (32 \,{\rm nm/s})^2 } \,.
    \label{eq:velo_can}
\end{equation}	              
\begin{equation}
  \Delta^{(3)} = \frac{\bigl[V([{\rm ADP}]^{(3)}) - V^E([{\rm ADP}]^{(3)})\bigr]^2}{\sigma_{V^E([{\rm ADP}]^{(3)})}^2}
      = \frac{\bigl[V([{\rm ADP}]^{(3)}) - 130 \,{\rm nm/s} \bigr]^2}{ (13 \,{\rm nm/s})^2 } \,.
\end{equation}
When the ATP concentration decreases the velocity $V^E$ is found to
decrease \cite{Yildiz-Forkey-McKinney-Ha-Goldman-Selvin2003,
Baker-Krementsova-Kennedy-Armstrong-Trybus-Warshaw2004pnas}. Choosing
$[{\rm ATP}]^{(2)}$ = 10 $\mu$M we construct a fourth term:
\begin{equation}
  \Delta^{(4)} = \frac{\bigl[V([{\rm ATP}]^{(2)}) - V^E([{\rm ATP}]^{(2)})\bigr]^2}
{\sigma_{V^E([{\rm ATP}]^{(2)})}^2} 
    = \frac{\bigl[V([{\rm ATP}]^{(2)}) - 75 \,{\rm nm/s} \bigr]^2}{ (10 \,{\rm nm/s})^2 } \,.
\end{equation}	              
It is found that changing [${\rm P_i}$] several millimolar does not
significantly change the velocity of myosin-V \cite{Mehta2001,
Baker-Krementsova-Kennedy-Armstrong-Trybus-Warshaw2004pnas}. Choosing
$[{\rm P_i}]^{(2)}$ = 40 mM, we add a fifth term using
measurements by \citet{Baker-Krementsova-Kennedy-Armstrong-Trybus-Warshaw2004pnas}:
\begin{equation}
  \Delta^{(5)} = \frac{\bigl[V([{\rm P_i}]^{(2)}) - V^E([{\rm P_i}]^{(2)})\bigr]^2}{\sigma_{V^E([{\rm P_i}]^{(2)})}^2} 
    = \frac{\bigl[V([{\rm P_i}]^{(2)}) -   440 \,{\rm nm/s}\bigr]^2}{ (44 \,{\rm nm/s})^2 } \,.
    \label{eq:velo_pi}
\end{equation}	              

\citet{Baker-Krementsova-Kennedy-Armstrong-Trybus-Warshaw2004pnas}
have performed a large number of experiments of the run length $L$ at
different nucleotide concentrations finding that the run length is
increased with a decrease in the concentration of either ADP or
ATP. We introduce four terms to the cost function based on run lengths
at different nucleotide concentrations (where $[{\rm ATP}]^{(3)}$ = 100 $\mu$M):
\begin{equation}
  \Delta^{(6)} = \frac{\bigl[L - L^E\bigr]^2}{\sigma_{L^E}^2} =
  \frac{\bigl[L - 800 \,{\rm nm} \bigr]^2}{ (150 \,{\rm nm})^2 } \,.
\end{equation}	              
\begin{equation}
  \Delta^{(7)} = \frac{\bigl[L([{\rm ADP}]^{(3)}) - L^E([{\rm ADP}]^{(3)})\bigr]^2}{\sigma_{L^E([{\rm ADP}]^{(3)})}^2} 
    = \frac{\bigl[L([{\rm ADP}]^{(3)}) - 400 \,{\rm nm}\bigr]^2}{ (150 \,{\rm nm})^2 } \,.
    \label{eq:runlength_can}
\end{equation}	              
\begin{equation}
  \Delta^{(8)} = \frac{\bigl[L([{\rm ATP}]^{(3)}) - L^E([{\rm ATP}]^{(3)})\bigr]^2}
	{\sigma_{L^E([{\rm ATP}]^{(3)})}^2} 
    = \frac{\bigl[L([{\rm ATP}]^{(3)}) - 1150 \,{\rm nm} \bigr]^2}{ (150 \,{\rm nm})^2 } \,.
\end{equation}	              
\begin{equation}
  \Delta^{(9)} = \frac{\bigl[L([{\rm P_i}]^{(2)}) - L^E([{\rm P_i}]^{(2)})\bigr]^2}
	{\sigma_{L^E([{\rm P_i}]^{(2)})}^2} 
    = \frac{\bigl[L([{\rm P_i}]^{(2)}) - 500 \,{\rm nm} \bigr]^2}{ (150 \,{\rm nm})^2 } \,.
    \label{eq:runl_pi}
\end{equation}	              
The velocity is found to be independent of external force up to
$f_{\rm ex}^{(2)}=0.75$ pN
\cite{Uemura-Higuchi-Olivares-DeLaCruz-Ishiwata2004,
Rief-Rock-Mehta-Mooseker-Cheney-Spudich2000, Mehta2001} and we include
an tenth term based on this:
\begin{equation}
  \Delta^{(10)} = \frac{\bigl[V(f_{\rm ex}^{(2)}, [{\rm ADP}]^{(2)}) 
  - V^E(f_{\rm ex}^{(2)}, [{\rm ADP}]^{(2)})\bigr]^2}{\sigma_{V^E(f_{\rm ex}^{(2)})}^2} 
    = \frac{\bigl[V(f_{\rm ex}^{(2)}) -   V([{\rm ADP}]^{(2)}) \bigr]^2}{ (50 \,{\rm nm/s})^2 } \,,
\end{equation}	              
where $V([{\rm ADP}]^{(2)})$ is the velocity in Eq.~\eqref{eq:velo_can}.  Similarly, it
was found that run length is fairly independent of external force
\cite{Clemen-Vilfan-Jaud-Zhang-Barmann-Rief2005}, giving the eleventh term:
\begin{equation}
  \Delta^{(11)} = \frac{\bigl[L(f_{\rm ex}^{(2)}, [{\rm ADP}]^{(2)}) 
  - L^E(f_{\rm ex}^{(2)}, [{\rm ADP}]^{(2)})\bigr]^2}{\sigma_{L^E(f_{\rm ex}^{(2)})}^2} 
    = \frac{\bigl[L(f_{\rm ex}^{(2)}) -   L([{\rm ADP}]^{(2)}) \bigr]^2}{ (150 \,{\rm nm})^2 } \,,
\end{equation}	              
where $L([{\rm ADP}]^{(2)}) = 400$ nm \cite{Baker-Krementsova-Kennedy-Armstrong-Trybus-Warshaw2004pnas}.
Also dwell time is found to be independent of external force up to $f_{\rm ex}^{(3)}=1$ pN 
\cite{Mehta-Rock-Rief-Spudich-Mooseker-Cheney1999, Uemura-Higuchi-Olivares-DeLaCruz-Ishiwata2004}
giving rise to the twelfth term:
\begin{equation}
  \Delta^{(12)} = \frac{\bigl[t_{\rm d}(f_{\rm ex}^{(3)}, [{\rm ADP}]^{(2)})
    - t_{\rm d}^E([{\rm ADP}]^{(2)})\bigr]^2}{\sigma_{t_{\rm d}^E(f_{\rm ex}^{(3)})}^2}
        = \frac{\bigl[t_{\rm d}(f_{\rm ex}^{(3)}) -   0.15\, s \bigr]^2}{ (0.1 \, s)^2 } \,.
\end{equation}
\citet{Rosenfeld-Sweeney2004} found that the release of ADP from the front head is at 
least 50 times slower than from the rear head, which gives rise to the thirteenth term:
\begin{equation}
  \Delta^{(13)} = \frac{1}{50^2} \bigl( \frac{J_{\rm fut} d}{V} \bigr)^2\,.
\end{equation}
where $J_{\rm fut}$ is the futile flux (Eq.~\eqref{eq:futflux}). The last four terms are restrictions on the
possible values of the energy jumps $\Delta G_i$, reflecting some inherent limits  
caused by strict limits on the energy
available in each sub-step of the chemical reaction.
Using measurements on S1 (see table~\ref{tab:delta_G}) as rough `target' values,
but allowing for a deviation from these values of $\sigma_{\Delta G_i}$ = 3 $k_{\rm B} T$ the last four terms are:
\begin{equation}
  \Delta^{(14-17)} = \frac{\bigl[\Delta G_i   - \Delta G_i^{\rm S1} \bigl]^2 }{\sigma_{\Delta G_i}^2} \,.
\end{equation}

The total cost function is simply defined to be the sum of all the
different cost terms:
\begin{equation}
  \Delta = \sum_{i=1}^{17} \Delta^{(i)} \,.
\end{equation}	              
All the terms in the cost function cannot be expected to constrain completely
independent properties of the model. However, with 17 differing constraints on
only nine free parameters it is encouraging, and perhaps not surprising, that the optimization reveals that the best solutions are grouped in the same region of parameter space.

% % % % % % % % % % % %  BIBLIOGRAPHY % % % % % % % % % % % 
  \bibliography{Ref}
% % % % % % % % % % % % % % % % % % % % % % % % % % % % % %

\clearpage
\begin{table}
  \centering
\begin{tabular}{c c c c c c c c c c}
$G^{\ddagger}_{2}$ & $G^{\ddagger}_{3}$ & $G^{\ddagger}_{4}$ & $G^{\ddagger}_{5}$ & $E_{\rm strain}$ & $\alpha E_{\rm strain}$
& $\Delta G_{2}$ & $\Delta G_{3}$ & $\Delta G_{4}$ & $\Delta G_{5}$
 \\
\hline \\
0.3 & 10.4 & 15.7  & 5.8  & 12.8  & 5.4 & 0.14  & 9.9 & $-10$ & 13.1 \\
\end{tabular}
  \caption{The estimated values (in units of $k_{\rm B}T$ ) of the free parameters in the model as given by
  the optimisation routine (see also Fig.~\ref{fig:EnergyLand}).}
     \label{tab:parameters}
\end{table}

\begin{table}
   \centering
\begin{tabular}{c c c c}
$\Delta G_{2}^{\rm S1}$ & $\Delta G_{3}^{\rm S1}$ & $\Delta G_{4}^{\rm S1}$ & $\Delta G_{5}^{\rm S1}$
 \\
 \hline\\
2 & 5.7 & $-7.7$ & 15.3 \\
\end{tabular}
  \caption{Energy jumps (in $k_{\rm B}T$ units) measured for the reaction of S1 with actin
  (taken from Table 14.2, Ref.~\cite{Howard_book} using Eq.~\ref{Eq:Ghyd}). }
     \label{tab:delta_G}
\end{table}

\clearpage
\section*{Figure Legends}
\subsubsection*{Figure~\ref{fig:ModelFig}.}
Sketch of the complete reaction network of the model. The Y-shaped
molecule is the myosin-V protein which walks on actin filaments. The
black actin monomers indicate the attachment sites spaced at $\simeq$
36 nm. The labels T, D, and ${\rm P_{i}}$ stand for ATP, ADP, and
inorganic phosphate respectively being bound to the head.
%%%% Fig. 2
\subsubsection*{Figure~\ref{fig:Cycle}.}
The main and futile cycles combined in one scheme showing all the
reactions paths between the 7 states in the complete model. See also
Fig.~\ref{fig:ModelFig}. The reaction rates are given by the corresponding equations in the text. Reaction
steps which release and bind ADP, ${\rm P}_{i}$, or ATP are indicated.
\subsubsection*{Figure~\ref{fig:Mech}.}
The mechanical movement of myosin-V takes place in two separate
steps. The first step, through a distance $d_W \simeq 25$ nm, is from the highly strained state 1 to state 2
where the internal strain balances the external force. When the
molecule diffuses to state 3, through a further distance $d_D \simeq 11$ nm, the internal strain increases to $b
E_{\rm strain}$.
%%%% Fig. 4
\subsubsection*{Figure~\ref{fig:EnergyLand}.}
The one dimensional energy landscape that we find for the walk of myosin-V in which the states in the model are indicated by the filled circles. Energy is measured in units of $k_B T$. The energy changes associated with the dashed transitions are $E_{\rm strain}$ and $b E_{\rm strain}$, being the energy barriers involved in moving away from state 2. Also shown is the rate limiting activation energy $G^{\ddagger}_4$ between state
4 and state 5. 
The generalised reaction coordinate $X$ can be thought of as measuring the progress around the main reaction cycle (Fig.~\ref{fig:ModelFig}). As such it reflects a combination of physical motion and the progress of biochemical reactions, according to the substep. The shape of the curve is somewhat arbitrary, but the peaks and
the troughs are at the correct energies determined by the optimal values ($G^{\ddagger}_{i}$, $E_{\rm strain}$, and $\Delta G_{i}$).
\subsubsection*{Figure~\ref{fig:Conc_Velo}.}
Predictions of velocity of myosin-V as a function of ADP, ATP, and
${\rm P_i}$ concentration that arise from our optimised (best) model with parameter values as shown in Table 1 (and used in all subsequent figures). In each case the other two reference concentration are taken from [ATP] = 1 mM, [ADP] = 0.1 $\mu$M, or [${\rm P_i}$] = 0.1 $\mu$M. The experimental 
data for varying [ATP] (squares) and [ADP] (circles) are from \citet{Baker-Krementsova-Kennedy-Armstrong-Trybus-Warshaw2004pnas}.
%%%% Fig. 6
\subsubsection*{Figure~\ref{fig:Fex_Velo}.}
The velocity as a function of force. The solid line (and circles) when
[ATP] = 1~mM and [ADP] = 200~$\mu$M, the dashed line (and squares) when
[ATP] = 1~mM and [ADP] = 1~$\mu$M, while for the dotted line (and
diamonds) [ATP] = 10~$\mu$M and [ADP] = 1~$\mu$M.  The model (the
lines) show similar trends to what is found experimentally (the
circles, squares, and diamonds)
\cite{Uemura-Higuchi-Olivares-DeLaCruz-Ishiwata2004}.
\subsubsection*{Figure~\ref{fig:DwellTime}.}
Dwell time for [ATP] = 2 mM (solid line/circles), [ATP] = 10 $\mu$M
(dotted line/squares), and [ATP] = 1 mM and [ADP] = 200 $\mu$M 
(dashed line/triangles). The experimental data are from
\citet{Mehta-Rock-Rief-Spudich-Mooseker-Cheney1999} (circles) and 
\citet{Uemura-Higuchi-Olivares-DeLaCruz-Ishiwata2004} (squares and triangles).
%%%% Fig. 8
\subsubsection*{Figure~\ref{fig:Fex_DutyRatio}.}
The duty ratio, $r_{\rm d}$, as a function of force for [ATP]= 1 mM, [ADP]=[${\rm P_i}$]=0.1 $\mu$M.
\subsubsection*{Figure~\ref{fig:ATP_RunL}.}
Run length $L$ for different concentrations of ATP when the ADP
concentration is equal to 1 mM (dotted--dashed line), 
100~$\mu$M (dashed line), and 10~$\mu$M (solid
line). At low ADP concentration, the run length becomes independent of
ATP concentration. The prediction of the model is compared with
experimental results at low ADP concentrations (circles)
\cite{Baker-Krementsova-Kennedy-Armstrong-Trybus-Warshaw2004pnas}.
%%%% Fig. 10
\subsubsection*{Figure~\ref{fig:Fex_RunL}.}
Run length $L$ for different strengths of the external force when
[ATP] = 1 mM and [ADP] = 200 $\mu$M. For negative external force the
run length has a non-monotonic behaviour, where it increases ten fold
before decreasing again.
\subsubsection*{Figure~\ref{fig:BackFrac}.}
The fraction of backward steps, $p_{\rm rev}/p_{\rm for}$, is
insignificant in the model until an external force of $\sim$2 pN is reached. The solid
line is for [ATP] = 2 mM and [ADP] = 200 $\mu$M, while the dashed line is
for a reduced ATP concentration of 100 $\mu$M.
%%%% Fig. 12
\subsubsection*{Figure~\ref{fig:Randomness}.}
The randomness ratio, $\rho$, as a function of force at different ATP
concentrations.
\subsubsection*{Figure~\ref{fig:Robustness}.}
The circle shows the unperturbed velocity and run length for [ATP]= 1 mM, [ADP]=[${\rm P_i}$]=0.1 $\mu$M and zero external force. The squares show the influence on run length and velocity of changes of
$\pm$5\% in each of the nine free parameters of the model (while
keeping the other parameters fixed). The run length is very sensitive
to changes in $G^{\ddagger}_4$. Also perturbing in $E_{\rm strain}$ gives
quite large change in run length. 
Large variation in velocity was only observed when perturbing the parameter $\Delta G_4$. 

\begin{figure}
  \centering
   \includegraphics[width=130mm]{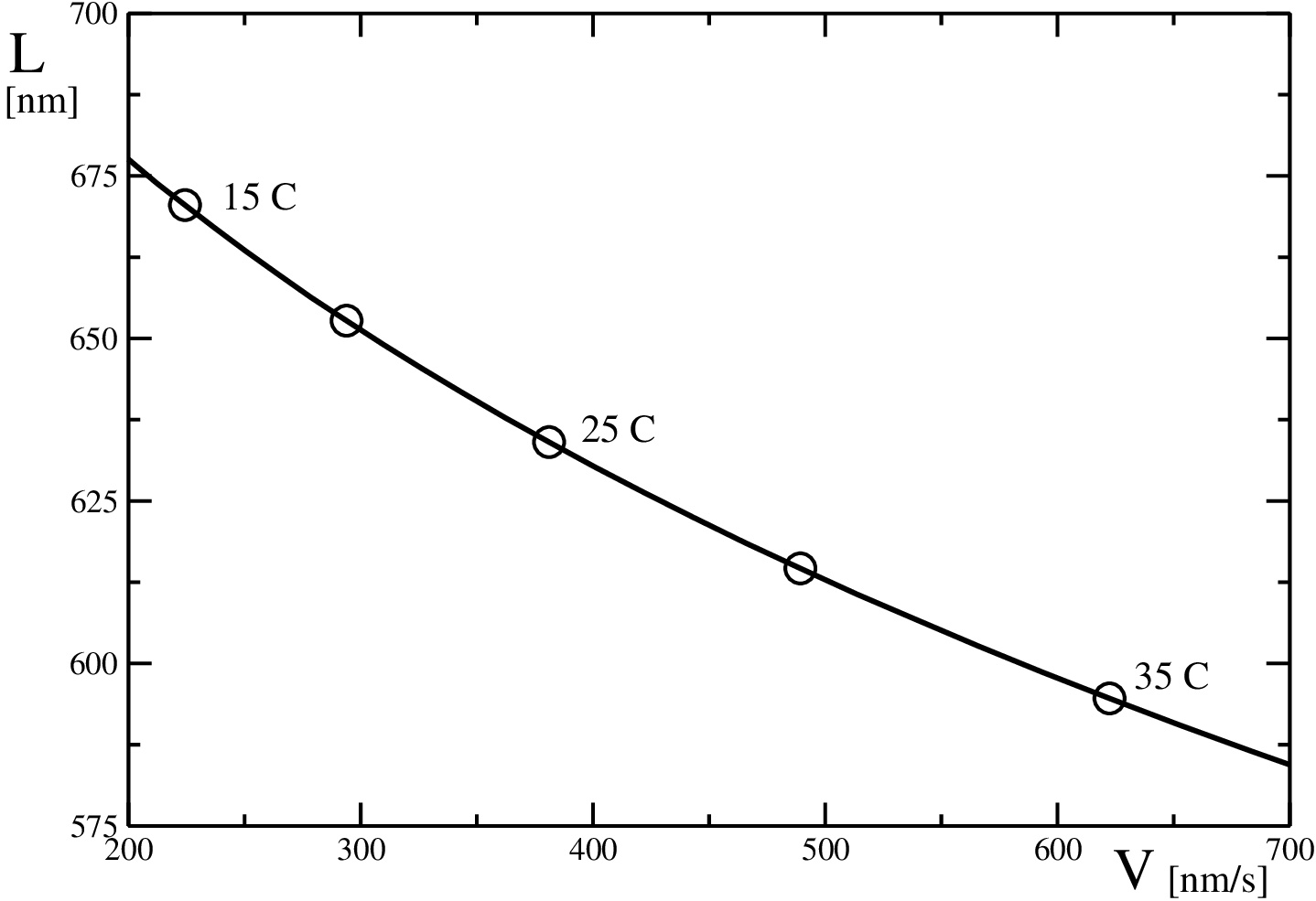}
   \caption{}
      \label{fig:Temp}
\end{figure}

%%%% Fig. 14
\subsubsection*{Figure~\ref{fig:Temp}.}
Temperature dependence of velocity and run length (when [ATP] = 1 mM
and [ADP] = 200 $\mu$M). The model predicts an increase in velocity with
temperature, but a decrease of the run length. The velocity is found
to be more sensitive to changes in temperature than the run length.

%%%% FIGURES and TABLES

\clearpage
\clearpage
\clearpage
\clearpage
\begin{figure}
\psfrag{E}[cc][cc]{\large E}
\psfrag{X}[cc][cc]{\large X}
\psfrag{[kT]}[cc][cc]{\normalsize ${\rm [k_BT]}$}
  \centering
   \includegraphics[width=130mm]{Fig4.eps}
   \caption{}
        \label{fig:EnergyLand}
\end{figure}
\clearpage
\clearpage
\clearpage
\clearpage
\clearpage
\clearpage
\clearpage
\clearpage
\clearpage
\clearpage
%

% - - - - - - - - - - - - - - - - - - - - - - - - - - - - - - %
            \end{document}